\documentclass[12pt]{article}
\usepackage{eurosym}
\usepackage{geometry}
\usepackage{amssymb}
\usepackage{amsmath,bm}
\usepackage{amsfonts}
\usepackage{graphicx,epsfig,lscape}
\usepackage{xcolor}
\usepackage{setspace}
\usepackage{url}
\usepackage{upgreek}
\usepackage{commath}
\usepackage{enumitem}
\usepackage{subfigure}
\usepackage{indentfirst}
\usepackage{xr}
\usepackage{hyperref}
\usepackage{natbib}
\usepackage{booktabs}
\usepackage{pifont}
\usepackage{cleveref}
\usepackage{titling}
\usepackage{multibib}
\usepackage{arydshln}
\allowdisplaybreaks
\newtheorem{theorem}{Theorem}

\newtheorem{assumption}{Assumption}

\newtheorem{example}{Example}

\newenvironment{proof}[1][Proof]{\noindent\textbf{#1.} }{\ \rule{0.5em}{0.5em}}
\begin{document}

\title{Orchestrating Organizational Politics: Baron and Ferejohn Meet Tullock
\thanks{
We thank Nageeb Ali, Jidong Chen, Yi-Chun Chen, Wouter Dessein, Ilwoo Hwang, Sam Jindani, Michihiro Kandori, Jin Li, Wolfgang Pesendorfer, Carlo Prato, Dmitry Ryvkin, Satoru Takahashi, Tianyang Xi, Huseyin Yildirim, Junjie Zhou, and seminar/conference participants at the 10th Annual Conference on Contests: Theory and Evidence, the 35th Stony Brook International Conference on Game Theory, the 7th World Congress of the Game Theory Society, the 2024 China Meeting on Game Theory and Its Applications, and the 2024 INFORMS Annual Meeting for helpful discussions, suggestions, and comments. Fu thanks the Singapore Ministry of Education Tier-1 Academic Research Fund (R-313-000-139-115) for financial support. Wu thanks the National Natural Science Foundation of China (Nos.\thinspace 72222002 and 72173002), the Wu Jiapei Foundation of the China Information Economics Society (No.\thinspace E21100383), and the Research Seed Fund of the School of Economics, Peking University, for financial support. Any errors are our own.} \\
}
\author{Qiang Fu\thanks{
Department of Strategy and Policy, National University of Singapore, 15 Kent
Ridge Drive, Singapore, 119245. Email: \href{mailto:bizfq@nus.edu.sg}{
bizfq@nus.edu.sg}.} \and Zenan Wu\thanks{
School of Economics, Sustainability Research Institute, Peking University,
Beijing, China, 100871. Email: \href{mailto:zenan@pku.edu.cn}{
zenan@pku.edu.cn}.} \and Yuxuan Zhu\thanks{
School of Economics, Peking University, Beijing, China, 100871. Email: \href{mailto:zhuyuxuan@pku.edu.cn}
{zhuyuxuan@pku.edu.cn}.}}
\date{November 8, 2024}
\maketitle

\begin{abstract}
\thispagestyle{empty} This paper examines the optimal organizational rules that govern the process of dividing a fixed surplus. The process is modeled as a sequential multilateral bargaining game with costly recognition. The
designer sets the voting rule---i.e., the minimum number of votes required to approve a proposal---and the mechanism for proposer recognition, which is modeled as a biased generalized lottery contest. We show that for diverse design objectives, the optimum can be achieved by a dictatorial voting rule, which simplifies the game into a standard biased contest model.

\begin{description}
\item[Keywords:] Multilateral Bargaining; Costly Recognition; Contest Design

\item[JEL Classification Codes:] C70; C78; D72.\newpage
\end{description}
\end{abstract}

\pagenumbering{Alph}

\pagenumbering{arabic} \setcounter{page}{1}

\section{Introduction}

Organizations---whether firms, academic institutions, political parties,
etc.---are political structures that \textquotedblleft operate by
distributing authority and setting a stage for the exercise of power\ 
\citep{zaleznik1970power}.\textquotedblright\ Organizational power grants
individuals preferred access to scarce resources or broader oversight of
vital activities (\citealp*{black2022simulated,pfeffer1993barriers}), which, in
turn, motivates efforts to acquire power and leverage it to influence key
decision-making within an organization. Such dynamics are commonplace and
inherent in organizational life. For instance, executives strive to ascend
the hierarchical ladders to positions that offer significant authority over
corporate agendas. Politicians campaign for electoral nominations or party leadership \citep{mattozzi2015mediocracy}. Members of
blockchain-technology-enabled decentralized autonomous organizations (DAOs)
devise and contribute proposals to mobilize community resources for funding
their projects.\footnote{
Readers are referred to \href{https://en.wikipedia.org/wiki/Decentralized\_autonomous\_organization}{https://en.wikipedia.org/wiki/Decentralized\_autonomous\_organization}.}

In this paper, we delve into the crafting of rules that govern this process
and regulate the interactions between individuals within an organization. We
identify two critical institutional elements that underpin these
organizational rules: (i) the organization's evaluation and promotion
mechanism, which selects key executives and allocates power and authority, and
(ii) the protocol for collective decision-making that either expands or
limits the executive's power to influence the distributive outcome. The
institutional setting defines how power is to be acquired, whom is to be
awarded the power, and the value placed on that power for each individual.
The choice of institutional elements addresses the ultimate challenge for
effective organization design, which calls for \textquotedblleft balancing
control mechanisms with incentives that drive performance\textquotedblright\
(\citealp{galbraith1973designing}), and speaks to the enduring debate over centralized vs.
decentralized power structure within organizations (\citealp*{mintzberg1973nature,kotter1985power,alonso2008does}).

For this purpose, we examine organizational interactions through the lens of
a multilateral sequential bargaining process (\citealp{baron1989bargaining}
). A pool of agents---e.g., business units inside a firm,
academic departments of a school, politicians within a political party, or
community members of a DAO---divide a fixed amount of resources. One agent
proposes a plan for resource sharing, and must secure a minimum number of
favorable votes from peers to approve and implement the plan. The
conventional wisdom of the literature on multilateral bargaining holds that
the proposer enjoys a disproportionately large share, which drives the
contest prior to the bargaining: Agents expend costly efforts to compete for the right to propose, which reflects the costly recognition mechanism \`{a} la 
\citet{yildirim2007proposal} and \citet{ali2015recognition}.

We endogenize the rules of this process. A designer sets (i) the bargaining
protocol and (ii) the recognition mechanism---i.e., the rules that govern the
selection of proposer. The former is defined by a $k$-majority rule, whereby $k
$ denotes the minimum number of favorable votes (including the proposer) required for approval. This
depicts the inclusiveness of the collective decision-making process and the
limit of the power or influence the proposer enjoys. The latter converts
agents' efforts into their recognition probabilities. By varying the
recognition mechanism, the designer can effectively bias the competition in
favor of certain contenders, and thereby tilt the playing field and reshape
agents' incentives. For instance, a preferred candidate in a company's
succession process is often assigned a significant position---e.g., president or
COO---that enhances their visibility to board members, and a politician
competing for party leadership can be endorsed by powerful party elites
(\citealp{cross2013party}). 

The governance of an organization could address diverse interests. We allow for a general objective that accommodates concerns about effort profiles and agents' recognition probability profile. The designer values agents' productive effort contributions, so the objective function (weakly) increases with each agent's effort.\footnote{The conventionally assumed objectives in the literature on optimal contest design---such as total effort maximization and the expected winner's effort maximization---are both special cases.} Consider, for instance, a political activist's services that contribute to the party's electoral influence and the quality of its platform. Similarly, a corporate executive's performance not only advances their career but also adds to the firm's value. The objective also accounts for concerns about the ex ante power distribution---i.e., the recognition probability profile. 

The game is intuitive by nature, but its analysis poses a technical
challenge. To vie for recognition, agents weigh their potential payoffs from
winning---i.e., being recognized---against those from losing. This payoff
differential effectively functions as the \emph{prize spread} that motivates
their efforts in the contest. In contrast to a standard contest with a prize
value $k\geq 2$, the prize spread in this game is endogenously determined. With a non-dictatorial voting rule, the winner
offers a subset of peers---namely, agents in his winning coalition---their
equilibrium continuation values in the dynamic bargaining process to secure
their votes; a loser, on the other hand, receives his equilibrium
continuation value if included in a winning coalition, or nothing if
excluded. The prize spreads---which depend on agents' continuation
values---motivate their efforts, while agents' efforts determine their
recognition probabilities, the formation of each agent's winning coalitions,
and, ultimately, their continuation values in equilibrium.

The endogeneity, together with agents' heterogeneity in contest
technologies, effort cost functions, and/or patience levels, complicates the
analysis and differentiates the game from traditional bargaining or contest
models. A change in either institutional element---the voting rule or the recognition
mechanism---triggers complex effects and precludes the possibility of
unambiguous comparative statics.

We establish equilibrium existence in the game in \Cref{Section-equilibrium}, although uniqueness is
elusive. Due to the aforementioned complications, the game does
not yield a closed-form equilibrium solution, which renders the usual implicit
programing approach ineffective. We elaborate on the economic nature of
these institutional variables in \Cref{Sec: Role of Voting Rule,Sec: Role of
Recognition Mechanism}. We develop a technique to identify the optimum: We
demonstrate that when the designer can set both the voting rule and the
recognition mechanism, the optimum always requires a \emph{dictatorial
voting rule} with $k=1$, although the specific form of recognition mechanism
depends on the particular environmental factors. That is, a proposal is
accepted with the consent of only the proposer, which results in an agent's capturing the entire surplus once recognized and receiving nothing
otherwise. The game thus reduces to a standard static contest. Notably, this
result holds even if the designer's sole objective is the fair distribution
of bargaining power ex ante. The universal optimality of the dictatorial
voting rule raises a further question: Suppose the designer can flexibly
adjust the recognition mechanisms. Does a less inclusive voting rule---i.e.,
a smaller $k$---always improve the value of the objective function? We
demonstrate that this does not hold in general. However, our analysis shows
that when agents are relatively impatient---i.e., when their patience levels
are limited by an upper bound---monotonicity is preserved. The intuition of
our results will be discussed in greater depth in \Cref{Section-optimal-design} after we formally present the
results.

Our analysis not only provides novel theoretical insights but also generates
significant policy and managerial implications. The results offer an
alternative perspective on the endogenous formation of centralized internal
power structures. We demonstrate that bargaining power can be optimally
allocated to a single individual, even when the organization prefers equal
distribution of recognition probabilities. Specifically, our insights
illuminate long-standing debates on the organization of political
parties, such as whether a political party subject to electoral
accountability requires intra-party democracy (IPD). The ascent
of blockchain-based DAOs provides
another relevant context, in which voting protocols and proposing eligibility
are critical for governance through smart contracts. These implications will
be further explored in \Cref{Sec: Implications}.

\paragraph{Link to the Literature}

Our paper adds to the literature on multilateral bargaining by providing a
comprehensive analysis to endogenize the bargaining protocol and the
proposer recognition mechanism.

An extensive body of literature has emerged from the canonical framework
established by \citet{baron1989bargaining} to explore the process of
distributive politics---e.g., \citet{merlo1995stochastic,merlo1998efficient}
; \citet*{duggan2000bargaining}; \citet{eraslan2002uniqueness}; 
\citet{eraslan2002majority,eraslan2017some}; \citet{diermeier2011legislative}
; \citet*{diermeier2015procedural,diermeier2016bargaining}; \citet*{ali2019predictability}; and 
\citet{evdokimov2023equality}. The majority of this literature assumes that
the proposer is exogenously and randomly selected from the agents.

A small but growing strand of the literature considers the selection of the
proposer to be an integral part of the political process, and examines the
endogenous formation of bargaining protocols. \citet{yildirim2007proposal}
models the process to select proposers as a contest in which agents exert
costly effort to gain power, and pioneers the integration of a contest model
(generalized Tullock contest) with a multilateral bargaining game to
endogenize the recognition mechanism. \citet{yildirim2010distribution}
compares total effort and distributive outcomes between persistent and
transitory recognition procedures, and \citet{ali2015recognition} models the
recognition process as an all-pay auction.

Our paper extends the effort to incorporate recognition mechanisms in a
holistic distribution process and models the recognition process as an
influencing competition. Our work is closely related to 
\citet{yildirim2007proposal}. Similar to \citeauthor{yildirim2007proposal},
we adopt a generalized Tullock contest, but we introduce heterogeneous
production technologies with fewer restrictions, as well as nonlinear effort
cost functions. \citeauthor{yildirim2007proposal} conducts comparative
statics of the prevailing voting rule for homogeneous agents and shows that a
more inclusive voting rule---i.e., a larger minimum number of required
votes---always decreases lower total effort. In contrast, we explore optimal
rule design in a setting that allows for a general design objective,
heterogeneous agents, and multiple design instruments (voting rule and
recognition mechanism). Agents' heterogeneity catalyzes complex effects
with varying voting rules or contest rules, which prevents standard
comparative static analysis and differentiates our game from conventional
bargaining or contest models.

Several papers examine the endogenous formation of a bargaining protocol
without using a contest approach. \citet*{
diermeier2015procedural,diermeier2016bargaining} employ a pre-bargaining
process to determine proposal power in bargaining over policy. In 
\citet{mckelvey1992seniority,mckelvey1993initial,muthoo2014seniority}; and 
\citet{eguia2015legislative}, recognition probability is determined by
seniority, which is endogenously voted on at the end of each session. 
\citet{kim2019recognition} assumes that current and past proposers are
excluded from the pool of eligible candidates when a round of bargaining
fails to reach consensus. \citet{jeon2022emergence} assume that an agent's
recognition probability and bargaining power depend on the previous
bargaining outcome in a dynamic legislative bargaining model, which leads to an
oligopolistic outcome as the result of an evolutionary process. \citet*{
agranov2020persistence} examine, both theoretically and experimentally, a
repeated multilateral bargaining model in which the agenda setter can retain
his power with the majoritarian support of other committee members.

Our paper is closely related to \citet{jeon2022emergence} and \citet*{ali2019predictability}, who
demonstrate that power concentration could arise in the equilibrium. The
former study attributes the endogenous formation of oligopoly to the
influence of past bargaining outcomes. The latter shows that the prevailing
information structure could lead to extreme power in terms of distributive
outcomes when the voting rule is not unanimous. Neither of them involves a contest of proposing
rights or an endogenously set voting rule, which is the focus of this paper.

Our paper is also naturally linked to the literature on contest design and,
particularly, that on optimally biased contests. We develop a technique
similar to that of \citet{fu2020optimal} and \citet*{fu2023bid}, who
characterize the optimum without explicitly solving for the equilibrium. Our
analysis complements these studies by embedding the contest in a
multilateral sequential bargaining framework, which generates an endogenous
prize spread.

\smallskip

The rest of the paper is structured as follows. \Cref{Section-model} sets
up the model and the design problem. \Cref{Section-equilibrium}
characterizes the equilibrium. \Cref{Section-optimal-design} solves the
optimal design problem and provides examples of the case with a single
instrument, and \Cref{Section-conclusion} concludes. Proofs and derivations for examples are collected in Appendices A and B, respectively.

\section{Model Setup}

\label{Section-model}

The game proceeds in two stages. In the first stage, a designer sets the rules that govern agents' subsequent interaction. In the second stage, a set of agents interact to divide a fixed sum of surplus, which is modeled as a multilateral sequential bargaining process with costly recognition, \`{a} la 
\citet{yildirim2007proposal,yildirim2010distribution} and 
\citet{ali2015recognition}.

\subsection{Multilateral Sequential Bargaining with Costly Recognition}

A set of $n\geq 2$ agents, indexed by $\mathcal{N}:=\{1,2,\dots ,n\}$,
decide how to divide a dollar. In each period $t=0,1,2,\dots $, one agent
(proposer) makes a proposal $\bm{s}_{t}\in \triangle
^{n-1}:=\{(s_{1,t},\dots ,s_{n,t}):0\leq s_{i,t}\leq 1,\sum_{i\in \mathcal{N}
}s_{i,t}=1\}$, where $s_{i,t}$ denotes the share of the dollar each agent $i\in \mathcal{N}$
is to receive under this proposal. Agents simultaneously vote in favor of or
against the proposal. We assume a \textquotedblleft $k$-majority
\textquotedblright\ voting rule---with $1\leq k\leq n$---for this sequential
bargaining process: The proposal is approved if at least $k$ agents accept
it (including the proposer). Specifically, $k=n$ implies a unanimous rule
wherein the proposal can be vetoed by any single dissident; $k=\lfloor {n}/{2
}\rfloor +1$ refers to a simple majority rule; with $k=1$, the proposer
dictates the decision process.

At the beginning of each period $t$, a contest takes place in which each
agent exerts an effort $x_{i,t}\geq 0$ to vie for the proposing right, which
incurs a cost $c_{i}(x_{i,t})$.

\paragraph{Proposer Recognition Mechanism}

For a given effort profile $\bm{x}_{t}:=(x_{1,t},\ldots ,x_{n,t})$, an agent 
$i$ is recognized as the proposer for period $t$ with a probability 
\begin{equation}
p_{i}(\bm{x}_{t})=\left\{ 
\begin{array}{ll}
\dfrac{\tilde{f}_{i}(x_{i,t})}{\sum_{j\in \mathcal{N}}\tilde{f}
_{j}(x_{j,t})},\  & \sum_{j\in \mathcal{N}}\tilde{f}_{j}(x_{j,t})>0, \\ 
\dfrac{1}{n},\  & \sum_{j\in \mathcal{N}}\tilde{f}_{j}(x_{j,t})=0,
\end{array}
\right.  \label{equation: defintion-p}
\end{equation}
where $\tilde{f}_{i}(\cdot )$ is called the impact function in the contest
literature; it converts one's effort into his effective output in the
competition, taking the form 
\begin{equation}
\tilde{f}_{i}(\cdot ):=\alpha _{i}f_{i}(\cdot )+\beta _{i}, \forall \,i\in 
\mathcal{N}.  \label{equation: tilde-f}
\end{equation}
The function $f_{i}(\cdot )$ describes agent $i$'s actual production
technology, while the multiplicative bias $\alpha _{i}\geq 0$ and additive
headstart $\beta _{i}\geq 0$ are set by the designer as one part of the rules
governing the contest for the proposing right. We will elaborate on the
details later.

\paragraph{Preferences and Payoffs}

Each agent is risk neutral and has a discount factor $\delta _{i}\in (0,1)$.
Agents differ in the degrees of their patience. If a proposal is approved in
period $\tau $, an agent $i$'s discounted payoff is 
\begin{equation*}
\Pi _{i}:=\delta _{i}^{\tau }s_{i,\tau }-\sum_{t=0}^{\tau }\delta
_{i}^{t}c_{i}(x_{i,t})\text{,}
\end{equation*}
where $s_{i,\tau }$ is the share he receives under the approved proposal and 
$c_{i}(x_{i,t})$ the effort cost incurred in each period $t\in \{0,\ldots
,\tau \}$.\footnote{
If no agreement is reached, agent $i$'s discounted payoff is $\Pi
_{i}=-\sum_{t=0}^{+\infty }\delta _{i}^{t}c_{i}(x_{i,t})$.}

\paragraph{Solution Concept}

The bargaining game with costly recognition can be described as $\big\langle(
\tilde{f}_{i}(\cdot ))_{i\in \mathcal{N}},(c_{i}(\cdot ))_{i\in \mathcal{N}},
\bm{\delta},k\big\rangle$, where $(\tilde{f}_{i}(\cdot ))_{i\in \mathcal{N}}$
denotes the set of impact functions, $(c_{i}(\cdot ))_{i\in \mathcal{N}}$
the set of effort cost functions, $\bm{\delta}:=(\delta _{1},\ldots ,\delta
_{n})$ the set of discounting factors, and $k$ the voting rule.

We assume that agents use stationary strategies whereby for each period $t$,
agents' period-$t$ actions are independent of the history (see 
\Cref{Theorem: equilibrium} for details of the strategies). We adopt the
solution concept of the stationary subgame perfect equilibrium (SSPE) and
drop the time subscript $t$ throughout. A strategy profile is an SSPE if it
is stationary and constitutes a subgame perfect equilibrium of the second-stage game.

We impose the following mild and standard regularity conditions to ensure
equilibrium existence:

\begin{assumption}
\label{Assumption: h_i} 

For each $i\in\mathcal{N}$, $f_{i}(\cdot)$ and $c_{i}(\cdot)$ are twice
differentiable in $(0,+\infty)$, satisfying $f_{i}(0) = 0$, $f_{i}^{\prime
}(\cdot) >0$, $f_{i}^{\prime \prime }(\cdot) \leq 0$, $c_{i}(0) = 0$, $
c_{i}^{\prime}(\cdot) > 0$, and $c_{i}^{\prime\prime}(\cdot) \geq 0$.
\end{assumption}

\subsection{Rule Design: Instruments and Objectives}

We now lay out the design problem.

\paragraph{Design Instruments}

The designer adjusts two structural elements of the bargaining process. She
sets the voting rule, which is implemented by choosing $k$, the minimum
number of favorable votes required for the proposal's approval. Meanwhile,
she can adjust the mechanism for proposer recognition (contest rules), which
determines the probability of each agent's recognition for every given effort
profile.

Recall that each agent $i$'s impact function $\tilde{f}_{i}(\cdot )$ is
given by \eqref{equation: tilde-f}. The designer imposes the multiplicative
weights $\bm{\alpha}:=(\alpha_1,\ldots,\alpha_n)\in \mathbb{R}_{+}^{n}\setminus \{(0,\ldots ,0)\}$
---which scale up or down one's output---and additive headstarts $\bm{\beta}
:=(\beta_1,\ldots,\beta_n)\in \mathbb{R}_{+}^{n}$. We can view $(\bm{\alpha},\bm{\beta})$ as nominal
scoring rules. Alternatively, they can be viewed as the organizational
resources assigned to agents that alter their productivity or influence
(see, e.g., \citealp{fu2022disclosure}).

Both multiplicative weights $\bm{\alpha}$ and additive headstarts $\bm{\beta}
$ are broadly adopted in modeling biased contests: \citet*{
epstein2011political} and \citet*{franke2014lottery}, for instance,
consider the former; 
\citet{konrad2002investment,siegel2009all,siegel2014asymmetric} and \citet*{
kirkegaard2012favoritism} focus on the latter; and \citet*{franke2018optimal}
and \citet{fu2020optimal} allow for both. It is noteworthy that $\bm{\alpha}$
and $\bm{\beta}$ play different roles in impacting the contest's
outcome: $\bm{\alpha}$ alter the marginal returns of agents' efforts, while $
\bm{\beta}$ directly add to their effective output regardless of their efforts.

\paragraph{Design Objectives}

As will be shown later in \Cref{Theorem: equilibrium}, there is no delay in
each SSPE, and thus agents exert effort at most once on the equilibrium
path. The designer chooses $(\bm{\alpha},\bm{\beta},k)$ to maximize an
objective function $\Lambda (\bm{x},\bm{p})$, where $\bm{x}:=(x_{1},\ldots ,x_{n})$ and $\bm{p}:=(p_{1},\ldots ,p_{n})$ denote the profiles of equilibrium efforts and agents' recognition
probabilities, respectively. The following regularity condition is imposed.

\begin{assumption}
\label{Assumption: Objective function} Fixing $\bm{p}$, $\Lambda (\bm{x},
\bm{p})$ weakly increases with $x_{i}$ for each $i\in \mathcal{N}$.
\end{assumption}

By \Cref{Assumption: Objective function}, we focus on the scenario in which
agents' efforts are productive and accrue to the designer's benefit.
Consider, for example, executives who enhance their performance to climb the
corporate ladder or party activists who contribute services to vie for leadership.

The objective function accommodates a diverse array of preferences.
Consider, for example, $\Lambda (\bm{x},\bm{p})=\sum_{i\in \mathcal{N}
}x_{i}-\lambda \sum_{i\in \mathcal{N}}\abs{p_{i}-\frac{1}{n}}$, with $
\lambda \geq 0$, which clearly satisfies 
\Cref{Assumption: Objective
function}. When $\lambda =0$, this objective boils down to maximizing
equilibrium total effort, which is conventionally assumed in the contest
design literature. When $\lambda >0$, the designer's payoff depends on the
profile of agents' recognition probabilities. The term $\sum_{i\in \mathcal{N
}}\abs{p_{i}-\frac{1}{n}}$---i.e., the mean absolute deviation of $\bm{p}$
---increases in the dispersion of $\bm{p}$. The function thus depicts a
preference for a more equitable distribution of recognition opportunities,
which compels the designer to set rules to reduce $\sum_{i\in \mathcal{N}}
\abs{p_{i}-\frac{1}{n}}$.\footnote{\citet{eraslan2017some} examine the
distributive implications of voting rules. They show that unanimity may
paradoxically lead to more unequal distributive outcome. It is noteworthy
that in our context, the designer's fairness concern refers to her
preference for ex ante distribution of bargaining power among agents---i.e.,
the recognition probability profile---instead of ex post distribution of the
surplus.}

Alternatively, consider $\Lambda (\bm{x},\bm{p})=\sum_{i\in \mathcal{N}
}p_{i}x_{i}$, which is the expected winner's effort. Maximizing the expected
winner's effort has gained increasing attention in the literature (e.g., 
\citealp{moldovanu2006contest,barbieri2019winners}). For instance, a firm
often views its succession race as a process to develop managerial talent;
the firm might benefit from the chosen successor's investment in their areas
of expertise, since the losers often pursue alternative career paths,
especially in high-profile public firms. For instance, James McNerney and Robert Nardelli joined 3M and Home Depot, respectively, after they lost
the race to succeed Jack Welch at General Electric.

\section{Equilibrium Existence and Characterization}

\label{Section-equilibrium}

We now characterize the equilibrium. Let $\bm{v}:=(v_1,\ldots,v_n)$
be the set of agents' equilibrium expected payoffs and consider
stage-undominated voting strategies, such that agents vote as if they were
pivotal. Suppose that an agent is not recognized as the proposer. He accepts
a proposal if his share exceeds the discounted continuation value---i.e., $
s_{i}\geq \delta _{i}v_{i}$---and rejects it otherwise. The proposer, in
contrast, needs to select $k-1$ agents to form the least costly winning
coalition and offers them their continuation values. His expected
vote-buying cost is
\begin{equation*}
w_{i}=\sum_{j\neq i}\psi _{ij}\delta _{j}v_{j},
\end{equation*}
where $\psi _{ij}$ gives the probability of agent $i$'s including $j$ in his
offer. For each $j\in \mathcal{N}$, we further define $\mu _{j}:=\sum_{i\neq
j}\psi _{ij}p_{i}$ as agent $j$'s probability of being included in others'
winning coalitions before a proposer is recognized.

For each agent $i\in\mathcal{N}$, the expected gross payoff conditional on being the
proposer is $1-w_{i}$ and that when not being selected is $\frac{\mu _{i}}{
1-p_{i}}\delta _{i}v_{i}$. His effort $x_{i}$ solves the maximization
problem on the right-hand side of the following Bellman equation: 
\begin{equation}
v_{i}=\max_{x_{i}\geq 0}\left\{ p_{i}(x_{i},\bm{x}
_{-i})(1-w_{i})+[1-p_{i}(x_{i},\bm{x}_{-i})]\times \frac{\mu _{i}}{
1-p_{i}(x_{i},\bm{x}_{-i})}\delta _{i}v_{i}-c_{i}(x_{i})\right\} ,
\label{equation:Bellman}
\end{equation}
which yields the following first-order condition:
\begin{equation}
\underbrace{c_{i}^{\prime }(x_{i})}_{\text{marginal cost of effort}}\geq 
\underbrace{\frac{\tilde{f}_{i}^{\prime }(x_{i})}{\tilde{f}_{i}(x_{i})}
\times p_{i}(1-p_{i})\times \overbrace{\left( 1-w_{i}-\frac{\mu _{i}}{1-p_{i}
}\delta _{i}v_{i}\right) }^{\text{effective prize spread}}}_{\text{marginal
benefit of effort}}.  \label{equation:foc}
\end{equation}

\Cref{equation:Bellman,equation:foc} depict the strategic nature of this
game. The term $1-w_{i}-\frac{\mu _{i}}{1-p_{i}}\delta _{i}v_{i}$ is the
payoff differential between winning the competition for recognition and
losing it---i.e., the prize spread that motivates efforts. However, the
prize spread is endogenously determined, since $w_{i}$, $p_{i}$, $\mu _{i}$,
and $v_{i}$ all depend on agents' effort profile $\bm{x}=(x_{1},\ldots
,x_{n})$ and vice versa. These nuances differentiate the model from a
standard contest with a fixed prize or a standard multilateral sequential
bargaining game, dismissing the regularity typically assumed in conventional
frameworks. Our analysis obtains the following.

\begin{theorem}
\label{Theorem: equilibrium} Suppose that \Cref{Assumption: h_i} holds. For
each game $\langle (\tilde{f}_{i}(\cdot ))_{i\in \mathcal{N}},(c_{i}(\cdot
))_{i\in \mathcal{N}},\bm{\delta},k\rangle $, there exists an SSPE
characterized by $(\bm{x},\bm{v})$ and $\{\psi _{ij}\}_{i\neq j}$. In the
equilibrium, each agent $i\in \mathcal{N}$ exerts effort $x_{i}$ in each
period. If selected as the proposer, he forms a winning coalition of $k-1$
agents such that agent $j$ is included with probability $\psi _{ij}$ and
offers the agent $\delta _{j}v_{j}$. Otherwise, he accepts a proposer's
offer if and only if his share is no less than $\delta _{i}v_{i}$. The
equilibrium is unique when $k=1$.
\end{theorem}

\Cref{Theorem: equilibrium} establishes equilibrium existence of the game,
which paves the way for optimal rule design. The setting of 
\citet{yildirim2007proposal} assumes $\tilde{f}_{i}(0)=0$, linear cost
function, and weakly decreasing elasticity $x_{i}\tilde{f}_{i}^{\prime
}(x_{i})/\tilde{f}_{i}(x_{i})$ for each $i\in \mathcal{N}$. We relax these
restrictions, allowing for headstarts $\beta _{i}$---which could lead to $
\tilde{f}_{i}(0)\neq 0$---nonlinear cost functions $c_{i}(\cdot )$, and
unrestricted elasticity conditions.

The equilibria might be nonunique, and a closed-form
equilibrium solution is in general unavailable in our context.\footnote{Fixing a recognition probability profile---i.e., fixing an effort profile---the literature on multilateral bargaining has noticed that there exist multiple equilibria that differ in $\{\psi_{ij}\}$, but they result in the same profile of $(\mu_{1},\ldots,\mu_{n})$. An additional layer of equilibrium multiplicity may arise within our context in the sense that agents' effort profile may differ across equilibria.} It is well
known that asymmetric contests cannot be solved in closed form when the
number of contestants exceeds two. The nuances caused by the endogenous
payoff structure entails further complications. This nullifies the usual
implicit programming approach to optimal design. We develop a technique in
line with \citet{fu2020optimal}, which enables us to characterize the
optimum without explicitly solving for the equilibrium.

\section{Optimal Design of Organizational Rules}

\label{Section-optimal-design}

We now explore the optimal organizational rules. We first examine the
respective nature of each set of structural elements---i.e., voting rule $k$
and recognition mechanism $(\bm{\alpha },\bm{\beta })$---and their respective
roles in shaping agents' incentives and behavior. We first demonstrate how the
implications differ from the conventional wisdom in the literature.
We then present the main results---i.e., the optimum when the designer has
full flexibility to adjust $(\bm{\alpha },\bm{\beta },k)$.

\subsection{Role of the Voting Rule}\label{Sec: Role of Voting Rule}

We begin with the voting rule $k$. Suppose that the bargaining process
implements a more inclusive voting rule, i.e., increasing $k$. It generates
an effect on agents' prize spreads, which we call the (direct) \emph{
prize effect}. A larger $k$ changes both his winning prize---i.e., $1-w_{i}$
---and losing prize, i.e., $\frac{\mu _{i}}{1-p_{i}}\delta _{i}v_{i}$. A
proposer has to buy more votes if he wins, and he needs to buy votes from a
different set of his peers; each peer would demand a different offer,
since their continuation values change: All of these change $w_{i}$. Further, a
losing candidate may expect a different payoff because he is more likely to
be included in some winning coalitions, while the minimum share he would
accept also varies with the change in his continuation value: These change $
\frac{\mu _{i}}{1-p_{i}}\delta _{i}v_{i}$ accordingly.

Overall, the effect is ambiguous. More importantly, this effect is
\emph{nonuniform} among asymmetric agents, which further leads to the (indirect) 
\emph{rebalancing effect}. Imagine $k=1$, such that all agents have a prize
spread of $1$ irrespective of their patience levels, since the proposer can expropriate all surplus without
rallying support from his peers and one ends up with nothing once he fails to
be recognized. Suppose that $k$ increases to $2$. Agents' patience 
$\delta _{i}$ now plays a role in determining prize spreads.
Ceteris paribus, the most patient agent is least likely to be included in a
winning coalition. His losing value, $\frac{\mu _{i}}{1-p_{i}}\delta
_{i}v_{i}$, tends to rise less than the others, causing a smaller decrease
in his prize spread than those of the others. The nonuniform changes in
agents' prize spreads tilt the playing field of the contest, which,
together with the heterogeneity in impact and cost functions, alter agents'
incentives indefinitely.

We construct an example to illustrate the subtlety. Assume that the designer
aims to maximize the total effort, i.e., with an objective function $\Lambda
(\bm{x},\bm{p})=\sum_{i\in \mathcal{N}}x_{i}$. Fixing a neutral recognition
mechanism $\bm{\alpha}=(1,\dots ,1)$ and $\bm{\beta}=(0,\dots ,0)$, we
explore the optimal voting rule.

\begin{example}
\label{Example: nonmonotone in k}
Suppose that $n = 4$, $f_i(x_i) = \eta_i x_i$, $c_i(x_i) = \eta_i x_i$, with $\bm{\eta} = (1, 0.2, 0.2, 0.2)$ and $\bm{\delta} = (0.1, 0.5, 0.5, 0.5)$. The recognition mechanism is fixed and required to be neutral, with $\bm{\alpha} = (1,1,1,1)$ and $\bm{\beta} = (0,0,0,0)$. The designer chooses $k \in \{1,2,3,4\}$ to maximize the total effort. The equilibria under different voting rules are depicted in \Cref{table-optimal-k}, which demonstrates that the total effort of the process is maximized by setting $k=2$.

\begin{table}[h!]
	\centering
	\begin{tabular}{c||c|c|c|c}
		\hline\hline
		& \quad $k=1$ \quad\quad& \quad $k=2$ \quad\quad & \quad $k=3$ \quad\quad & \quad $k=4$ \quad\quad \\
		\hline
		Winning probability of agent 1   & 0.2500    & 0.2322  & 0.2421  & 0.2500    \\
		\hline
		Winning probability of agents 2-4 & 0.2500    & 0.2559  & 0.2526  & 0.2500    \\
		\hline
		Equilibrium effort of agent 1   & 0.1875  & 0.1711  & 0.1656  & 0.1570  \\
		\hline
		Equilibrium efforts of agents 2-4  & 0.9375  & 0.9433  & 0.8641  & 0.7849  \\
		\hline
		Total effort                     & 3.0000       & 3.0011  & 2.7578  & 2.5116  \\
		\hline
	\end{tabular}
	\caption{Equilibrium Outcomes in \Cref{Example: nonmonotone in k}.}
	\label{table-optimal-k}
\end{table}
\vspace{-10pt}
\end{example}

\vspace{-10pt}

The results are summarized in \Cref{table-optimal-k}. There are two types of
agents: $1$ impatient agent and $3$ patient agents. When $k=1$,
heterogeneity in effort cost and that in impact function perfectly offset
each other and all agents win with equal probability. Each agent needs to
pay the vote-buying cost when $k$ increases to $2$ and thus the prize effect
arises, which tends to reduce the prize spread and the equilibrium effort.
The three patient agents have higher levels of patience and are less likely
to be included in other agents' winning coalition, since high patience
elevates their continuation value and therefore others' costs of buying
their votes. Conversely, the impatient agent is always included in the
winning coalition, which increases his losing value. As a result, patient
agents have a larger prize spread and therefore a stronger prize incentive.
The nonuniform changes in prize spreads alter the balance of the playing
field and lead to the rebalancing effect, which increases patient agents'
winning probability, intensifies their competition, and tends to increase
their equilibrium effort. In this example, the rebalancing effect dominates
the opposing prize effect for the three patient agents. Fixing a neutral
recognition mechanism, the total effort is nonmonotone in $k$, being
maximized by $k=2$.

This observation sharply contrasts with the result of \citet{yildirim2007proposal}. With
symmetric agents, \citeauthor{yildirim2007proposal} shows that total effort strictly decreases
with $k$. Intuitively, with symmetric agents, an increase in $k$ decreases agents' prize
spreads and weakens their incentives, while the rebalancing effect is
entirely muted due to symmetry. As shown in \Cref{table-optimal-k}, for
fixed $(\bm{\alpha },\bm{\beta })$, the value of the objective function
is nonmonotone with respect to $k$ and no explicit prediction can be
obtained in general in the case in which agents are heterogeneous.

\subsection{Role of the Recognition Mechanism}

\label{Sec: Role of Recognition Mechanism}

We now examine the role played by the recognition mechanism. The designer sets $
(\bm{\alpha },\bm{\beta })$, while fixing the voting rule $k$, to
maximize the objective function
\begin{equation}
\Lambda (\bm{x},\bm{p})=\sum_{i\in \mathcal{N}}x_{i}-\lambda \sum_{i\in 
\mathcal{N}}\,\abs{p_{i} - \frac{1}{n}}\text{, with }\lambda >0.
\label{equation: objective function}
\end{equation}

\begin{example}
\label{example: optimal_beta>0} Suppose that $n=3$, $k=2$, $f_{i}(x_{i})$ $=$
$x_{i}$, and $c_{i}(x_{i})=c_{i}x_{i}$ with $(c_{1},c_{2},c_{3})=(1,1,c)$.
Let $(\delta _{1},\delta _{2},\delta _{3})=(\frac{3}{8},\frac{1}{2},\frac{12
}{13})$. Assume that $\lambda $ is sufficiently large and $c$ is
sufficiently small, with $\lambda \gg \frac{1}{c}\gg 1$.

The objective function can be maximized by a recognition mechanism with $
\bm{\alpha}^{\ast }=(\frac{62Y}{35},\frac{62Y}{37},\frac{62Yc}{39})$ and $
\bm{\beta}^{\ast }=(0,\frac{17Y}{222},0)$, where $Y>0$ is an arbitrary
positive constant. The game yields an equilibrium outcome of $\bm{x}=(\frac{
70}{372},\frac{57}{372},\frac{78}{372c})$ and $\bm{p}=(\frac{1}{3},\frac{1}{3
},\frac{1}{3})$. The designer's payoff is $\Lambda =\frac{127}{372}+\frac{78
}{372c}$. 
\begin{table}[th]
\begin{center}
\begin{tabular}{@{}l||c||c||c}
\hline\hline
& \qquad Agent 1 \qquad\qquad & \qquad Agent 2 \qquad\qquad & \qquad Agent 3
\qquad\qquad \\ \hline
\, Equilibrium efforts & ${70}/{372}$ & ${57}/{372}$ & ${78}/{(372c)}$ \\ 
\hline
\, Winning probability & ${1}/{3}$ & ${1}/{3}$ & ${1}/{3}$ \\ \hline
\, Equilibrium payoff & ${56}/{372}$ & ${72}/{372}$ & ${39}/{372}$ \\ \hline
\, Winning coalition & $\{1,2\}$ & $\{1,2\}$ & $\{1,3\}$ \\ \hline
\end{tabular}
\end{center}
\par

\vspace{-10pt}
\caption{Equilibrium Outcomes in \Cref{example: optimal_beta>0}.}
\label{table-equilibrium-winning-coalition}
\end{table}
\end{example}
\vspace{-10pt}

Notably, the designer awards a positive headstart to agent 2. This stands in contrast
to findings in the literature on contest design. 
\citet{fu2020optimal}, for instance, formally establish the suboptimality of
a headstart, and show that adjusting $\bm{\alpha }$ suffices to achieve
the optimum. The contrast unveils how the endogenous prize structure
differentiates the game from a standard static contest.

With $c_{1}=c_{2}>c_{3}$ and $\delta _{1}<\delta _{2}<\delta_{3} $, agent $3$ is
ex ante the strongest contender, followed by agent $2$, then agent $1$. The
designer would benefit if agent $3$ can be sufficiently incentivized given
his low effort cost, which requires a larger prize spread for the agent. For
this purpose, the designer can seek to reduce agent $1$'s continuation
value, which decreases agent 3's vote-buying cost---i.e., $w_{3}$---given
that by \Cref{table-equilibrium-winning-coalition}, agent 3 would include
agent $1$ in his winning coalition. 

Further, by \Cref{table-equilibrium-winning-coalition}, agent $1$ would buy
agent $2$'s vote upon being the proposer. The designer can increase agent $2$
's continuation value to render agent $1$ worse off, which can be achieved by
awarding agent $2$ either a headstart $\beta _{2}>0$ or a larger $\alpha _{2}
$. The former is more effective in this context: Both increase agent 2's
recognition probabilities and improve his payoffs. However, a larger $\alpha
_{2}$ increases the marginal benefit of effort, which promotes his effort
supply; effort is costly and, in turn, reduces agent 2's payoff, (partially)
offsetting the payoff-improving effect of a larger $\alpha _{2}$.

These subtleties, as the artifact of the dynamic bargaining process, are
absent in a simple static contest. The prize is fixed in a standard contest,
so the contest rule $(\bm{\alpha },\bm{\beta })$ only rebalances the
playing field and does not alter agents' prize incentives. Multiplicative
biases $\bm{\alpha 
}$ can more effectively motivate efforts due to their direct impact on the
marginal benefits of efforts, rendering headstarts $\bm{\beta}$ redundant. In
contrast, the prize spreads in our context, when $k\geq 2$, endogenously
depend on agents' equilibrium efforts; so a change in the contest rule
catalyzes not only a (direct) rebalancing effect but also an (indirect)
prize effect. In this particular example, varying $\bm{\beta }$ creates an
opportunity to exploit the endogenous payoff structure of the game.

In summary, due to the complexity, no general comparative statics can be obtained with
respect to the recognition mechanism $(\bm{\alpha },\bm{\beta })$
under fixed $k$.

\subsection{Main Result}

We now explore the general optimization problem that allows the designer to
set $k$ and $(\bm{\alpha },\bm{\beta })$ altogether to maximize the objective
function $\Lambda (\bm{x},\bm{p})$. Despite the complexity that arises when
either $k$ or $(\bm{\alpha} ,\bm{\beta} )$ varies alone, an optimum exists with
unambiguous implications. Our analysis concludes the following.

\begin{theorem}
\label{Theorem: optimal} Suppose that Assumptions \ref{Assumption: h_i} and 
\ref{Assumption: Objective function} hold. When the designer can flexibly
choose $(\bm{\alpha},\bm{\beta},k)$, the optimum involves a dictatorial
voting rule ($k=1$) and zero headstart ($\bm{\beta}=\bm{0}$).
\end{theorem}

By \Cref{Theorem: optimal}, a dictatorial voting rule always emerges in the
optimum, although the specific form of the associated recognition mechanism $
(\bm{\alpha },\bm{\beta })$ depends on the particular context. The
proposer does not need a winning coalition and relinquish his share. As a
result, each agent has a fixed prize spread of $1$ in the optimum, and their
patience levels do not affect an equilibrium outcome. It is noteworthy that the
optimum calls for ex post concentration of bargaining power, even if she
cares about an even distribution of recognition opportunity ex ante.

The logic of these results can be interpreted in light of the interactions
between the prize and the rebalancing effects postulated earlier. A dictatorial
voting rule ($k=1$) generates a maximized prize spread, since both $w_{i}$
and $\frac{\mu _{i}}{1-p_{i}}\delta _{i}v_{i}$ are zero. This provides the
largest prize incentive to agents and tempts them to strive for
recognition. The designer can then adjust the recognition mechanism $(
\bm{\alpha },\bm{\beta })$ to optimally tilt the playing field if
necessary. It is noteworthy that the ambiguous indirect prize effect caused
by a change in $(\bm{\alpha },\bm{\beta })$ is entirely muted because the
prize spread is fixed when $k=1$ and does not depend on agents' continuation
values. The findings from the contest literature (
\citealp{fu2020optimal}) can be reinstated: The designer can induce any
profile of equilibrium winning odds by adjusting $\bm{\alpha }$, and
additive headstarts $\bm{\beta }$ are redundant.

In summary, the joint design achieves the optimum, and the two sets of
instruments play distinct roles: The voting rule---with $k=1$---generates
the maximum and a fixed prize spread, while the multiplicative biases $
\bm{\alpha }$ optimally exploit agents' heterogeneity in terms of their innate
abilities and sets the optimal competitive balance.

\subsection{Extensions, Discussions, and Implications}

In what follows, we first present further discussions that explore the limit
of our analysis, then elaborate on the implications of our results for
organizational design.

\subsubsection{Further Analysis}

\Cref{Theorem: optimal} establishes the general optimality of a dictatorial voting rule,
although the specific form of the associated recognition mechanism depends
on the particular environment. This observation naturally inspires the
following question regarding the general effect of varying $k$: Despite the
unavailability of the comparative statics of $k$ when the voting rule
changes alone, as shown in \Cref{Example: nonmonotone in k}, does a less inclusive voting rule---i.e., a smaller $k$---necessarily improve the value of the objective function \eqref{equation: objective function} when the
designer can adjust the recognition mechanism optimally for every given $k$?

The following example demonstrates that this conjecture does not hold in
general.

\begin{example}\label{Example: non-monotonicity}

Suppose $n=7$ and $\delta _{i}=0.999$
for all agents $i\in\mathcal{N}$. Each agent has a production technology $
f_{i}(x_{i})=x_{i}$. We construct the following vectors: $
\bm{\tilde{p}}:=(0.005,0.005,0.005,0.1,0.1,\break 0.1,0.685)$, $
r=839.9\times \tilde{p}_{7}(1-\tilde{p}_{7})$, and $\bm{\tilde{x}}
=(0.0037,0.0037,0.0037,0.0144,0.0144,0.0144,\break0.0001^{\frac{1}{r}})$. Further, agents' effort cost functions take the following form: 
\begin{equation*}
c_{i}(x_{i})=\left\{ 
\begin{array}{ll}
x_{i}, & x_{i}\leq \tilde{x}_{i}\text{ and }i\leq 6, \\ 
x_{i}^{r}, & x_{i}\leq \tilde{x}_{i}\text{ and }i=7, \\ 
\tilde{x}_{i}+\gamma (x_{i}-\tilde{x}_{i}), & x_{i}>\tilde{x}_{i}\text{ and }
i\leq 6, \\ 
\tilde{x}_{i}^{r}+\gamma (x_{i}-\tilde{x}_{i}), & x_{i}>\tilde{x}_{i}\text{
and }i=7,
\end{array}
\right.
\end{equation*}
where $\gamma $ is a sufficiently large constant. Assume an objective function \eqref{equation: objective function}---i.e., $\Lambda =\sum_{i\in \mathcal{N}}x_{i}-\lambda \sum_{i\in \mathcal{N}} \abs{p_{i}-\tilde{p}_{i}}$---with a sufficiently large $\lambda $. The designer can freely set $(\bm{\alpha },\bm{\beta })$. It can be verified that setting $k$ to either $5$ or $1$ maximizes the objective function, while $k=4$ is suboptimal, which indicates the nonmonotonicity of the designer's payoff with respect to $k$.
\end{example}

\Cref{Example: non-monotonicity} sheds light on the nuances of this game.
On the one hand, increasing $k$ requires a proposer to buy more
votes, which, ceteris paribus, directly reduces each agent's winning prize
and effort incentive. On the other hand, the cost for each vote may also
change due to the altered dynamics involved in the bargaining process; this
indirectly affects agents' winning prizes and could either increase or
decrease them. By \Cref{Example: non-monotonicity}, the cost of an
individual agent's vote can be reduced when the voting rule becomes more
inclusive; so the latter indirect effect may dominate the former
direct effect, which results in the nonmonotonicity of the designer's payoff
with respect to $k$.

Our analysis further obtains the following.

\begin{theorem}\label{Thorem: monotonicity}
Suppose that $\delta _{i}\leq \frac{1}{2}$ for each $i\in \mathcal{N}$ and
Assumptions \ref{Assumption: h_i} and \ref{Assumption: Objective function}
hold. If the designer can flexibly adjust $(\bm{\alpha},\bm{\beta})$, the
objective function $\Lambda $ is weakly decreasing in $k$.
\end{theorem}

\Cref{Thorem: monotonicity} establishes the general monotonicity of $k$ with
additional restrictions on $\delta _{i}$. That is, the designer prefers a
less inclusive voting rule when agents are relatively impatient---i.e., when $
\delta _{i}$ is bounded from above by $1/2$. The result is intuitive.
When agents are relatively impatient, the abovementioned nuanced indirect
effect triggered by changes in $k$ and $(\bm{\alpha },\bm{\beta })$ are
limited. The direct effect of a rising $k$---i.e., elevating vote-buying
cost and reducing prize spread---dominates the trade-off and leads to our
prediction.

\subsubsection{Implications}\label{Sec: Implications}

Our analysis provides valuable insights for designing organizational rules
that guide the internal processes of distributing power and allocating
resources. For instance, our results could inform renewed debates on
intra-party democracy (IPD)---i.e., a party's practice of selecting leaders,
making key decisions, and deploying resources (\citealp{cross2013party,poguntke2016party}). Strong advocacy exists for reforms to democratize
the internal structures of political parties and address a wide array of
concerns, such as restoring public trust, promoting inclusion and openness,
and rebuilding the democratic link between citizens and governments (\citealp*{dalton2011political,scarrow1999parties,scarrow2014beyond}). As documented by \citet{cross2013party}, parties' practices differ substantially in reality. For example, party leaders in the UK and Belgium were often chosen by party
elites through acclamation, even in leadership contests with broad
selectorates or full member votes, whereas Canada sets a low threshold for
access to the contest by putting candidates on an equal footing. Parties also
vary significantly in terms of the inclusiveness of their decision-making
procedures and the authority granted to leaders (\citealp{poguntke2016party}).

The practice of IPD could be a double-edged sword, yielding both positive
and negative effects, especially given the external accountability pressure
imposed by electoral competition (\citealp{cross2013challenges}). A wealth of
scholarly effort has been dedicated to the debate on IPD in political
science literature, primarily addressing its normative concerns. The
economics literature, however, remains relatively silent in terms of providing formal
analysis regarding the ramifications of power distribution or internal party
organization. \citet{dewan2016defense} provide a rationale for faction
formation within a party, and demonstrate that factionalism facilitates information
aggregation and empowers moderate politicians. \citet{caillaud2002parties}
demonstrate how competition between factions for candidate selection can be
managed for electoral success. \citet*{crutzen2010party}
portray a context in which two parties choose their own internal
structures---i.e., the competitiveness of candidate selection
mechanisms---and identify critical effects of organizational formats on
electoral competition. Our paper focuses on a different context and a
different set of instruments to streamline the internal distributive
process, and thereby complements this strand of the literature.

As another example, our analysis could yield insights for the burgeoning
experimentation with blockchain-based decentralized autonomous organizations
(DAOs). A DAO is an organization in which members propose projects and vote
to decide whether a project can be funded by community resources. Although
the execution of organizational decisions is automated through smart
contracts based on blockchain-enabled computer programs, the prescribed
procedures (voting and funding) are subject to the choice of the DAO's
initiating institution.\footnote{
Readers are referred to
\href{https://en.wikipedia.org/wiki/Decentralized\_autonomous\_organization/\#cite\_note-17}{https://en.wikipedia.org/wiki/Decentralized\_autonomous\_organization\#cite\_note-17}.}

The governance of DAOs has sparked extensive debates on both equity and
efficiency grounds (\citealp*{zhao2022task}). The
organizational rules of a DAO typically mainly consist of two elements: (i)
the threshold for passing a proposal---e.g., the number of favorable votes
required and/or the length of the voting window---and (ii) the mechanism for
filtering or selecting proposals for voting. This scenario provides a particularly
relevant context for our analysis. According to our results, the optimum
reduces the process to a proposal selection mechanism while setting a
minimum threshold---i.e., leaving decision rights exclusively to the
proposer. This holds even if the design objective involves concerns about
the even distribution of decision power ex ante. Although operating a DAO
and implementing its rules could present various other challenges beyond the
scope of our model, the analysis offers a useful perspective.

\section{Concluding Remarks}

\label{Section-conclusion} In this paper, we model organizational politics
as a sequential multilateral bargaining game with costly recognition, in
which a proposer suggests a plan to divide a fixed amount of resources and
the proposer is determined through a contest. We explore the optimal
organizational rules, with a designer who deploys two sets of design
instruments: (i) the voting rule that governs how proposals are accepted or
rejected and (ii) the recognition mechanism that determines how the
proposer is selected based on agents' productive efforts. We demonstrate that when the
designer can choose both structural elements, the optimum always involves a dictatorial rule that maximizes a general objective function.
This simplifies the bargaining game with costly recognition and condenses it
into a standard contest.

\bibliographystyle{aernoboldcomma}
\bibliography{bargaining}

\clearpage

\section*{Appendix A: Proofs}

\label{Section: Appendix}

\begin{description}
\item[Proof of \Cref{Theorem: equilibrium}] 
\end{description}

\begin{proof}
We first characterize the SSPE assuming its existence, then prove
equilibrium existence.

\paragraph{Equilibrium Characterization}

Denote by $V^{\Delta }$ the $k$-th lowest continuation value. Let $\mathcal{
\ N}_{1}:=\{i\in \mathcal{N}:\delta _{i}v_{i}<V^{\Delta }\}$, $\mathcal{N}
_{2}:=\{i\in \mathcal{N}:\delta _{i}v_{i}=V^{\Delta }\}$, and $\mathcal{N}
_{3}:=\{i\in \mathcal{N}:\delta _{i}v_{i}>V^{\Delta }\}$. Evidently, agent $
i $, when becoming the proposer, buys out the votes of the cheapest
``winning coalition''---i.e., $\mathcal{N}_{1}$ and a subset of $\mathcal{N}
_{2}$, from which we can conclude 
\begin{equation}  \label{def-mu}
{\psi _{ij}\left\{ 
\begin{array}{ll}
=1, & j\in \mathcal{N}_{1}, \\ 
\in \lbrack 0,1], & j\in \mathcal{N}_{2}, \\ 
0, & j\in \mathcal{N}_{3},
\end{array}
\right. \text{and } \mu _{i}\left\{ 
\begin{array}{ll}
=1-p_i, & i\in \mathcal{N}_{1}, \\ 
\in [0,1-p_{i}], & i\in \mathcal{N}_{2}, \\ 
= 0, & i\in \mathcal{N}_{3}.
\end{array}
\right.}
\end{equation}

Define 
\begin{equation}  \label{equation: definition VL}
V_{L}:=1-\sum_{j\in \mathcal{N}_{1}}\delta _{j}v_{j}-\left(k-\abs{\mathcal{N}_{1}}\right)V^{\Delta }.
\end{equation}
Agent $i$'s expected cost is then 
\begin{equation*}
w_{i}=\left\{ 
\begin{array}{ll}
1-V_{L}-\delta _{i}v_{i}, & i\in \mathcal{N}_{1}, \vspace{5pt} \\ 
1-V_{L}-V^{\Delta }, & \text{otherwise.}
\end{array}
\right.
\end{equation*}

The effective prize spread $1-w_{i}-\frac{\mu _{i}}{1-p_{i}}\delta _{i}v_{i}$
in \eqref{equation:foc} can be expressed as
\begin{equation}  \label{temp-proof-1}
1-w_{i}-\frac{\mu _{i}}{1-p_{i}}\delta _{i}v_{i}=V_{L}+\frac{1-\mu
_{i}-p_{i} }{1-p_{i}}V^{\Delta }=\left\{ 
\begin{array}{ll}
V_{L},\  & i\in \mathcal{N}_{1}, \\ 
V_{L}+\frac{1-p_{i}-\mu _{i}}{1-p_{i}}V^{\Delta },\  & i\in \mathcal{N}_{2},
\\ 
V_{L}+V^{\Delta },\  & i\in \mathcal{N}_{3}.
\end{array}
\right.
\end{equation}

We are ready to lay out the conditions for equilibrium characterization. An
SSPE can be characterized by $(\bm{x},\bm{v},\bm{p},\bm{\mu},V_L,V^{\Delta})$
. Combining \eqref{equation:foc} and \eqref{temp-proof-1} yields
\begin{equation}  \label{equation: equilibrium-condition-1}
\frac{c_{i}^{\prime }(x_{i})\tilde{f}_{i}(x_{i})}{\tilde{f}_{i}^{\prime
}(x_{i})}\geq p_{i}(1-p_{i})\left( V_{L}+\frac{(1-p_{i}-\mu _{i})V^{\Delta } 
}{1-p_{i}}\right).
\end{equation}
Next, consider the expected payoff $v_{i}$. By \eqref{equation:Bellman}, we
have

\begin{equation}
v_{i}=p_{i}(1-w_{i})+\mu _{i}\delta _{i}v_{i}-c_{i}(x_{i})= \left\{ 
\begin{array}{ll}
\frac{1}{1-\delta _{i}}\left(p_{i}V_{L}-c_{i}(x_{i})\right), & i\in \mathcal{
N}_{1}, \\ 
\frac{V^{\Delta }}{\delta _{i}}, & i\in \mathcal{N}_{2}, \\ 
p_{i}(V_{L}+V^{\Delta })-c_{i}(x_{i}), & i\in \mathcal{N}_{3}.
\end{array}
\right.  \label{equation: equilibrium-condition-3}
\end{equation}

Combining \eqref{equation:Bellman}, \eqref{def-mu}, and 
\eqref{equation:
equilibrium-condition-3} yields 
\begin{equation}
\mu _{i}\left\{ 
\begin{array}{ll}
=1-p_{i}, & i\in \mathcal{N}_{1}, \\ 
\in \lbrack 0,1-p_{i}]\text{ solves }\frac{V^{\Delta }}{\delta _{i}}
=p_{i}V_{L}+(\mu _{i}+p_{i})V^{\Delta }-c_{i}(x_{i}), & i\in \mathcal{N}_{2},
\\ 
=0, & i\in \mathcal{N}_{3}.
\end{array}
\right.  \label{equation: equilibrium-condition-4}
\end{equation}
Each agent chooses exactly $k-1$ agents in his winning coalition---i.e., $
\sum_{j\not= i}\psi _{ij}=k-1,\ \forall i\in \mathcal{N}$.
Therefore, 
\begin{equation}
\sum_{i\in \mathcal{N}}\mu _{i}=\sum_{i\in \mathcal{N}}\sum_{j\not= i}\psi _{ji}p_{j}=\sum_{j\in \mathcal{N}}p_{j}\sum_{i\not= j}\psi_{ji}=\sum_{j\in \mathcal{N}}(k-1)p_{j}=k-1.
\label{equation: equilibrium-condition-5}
\end{equation}
Last, \eqref{equation: definition VL} can be rewritten as
\begin{equation}
V_{L}+\sum_{i\in \mathcal{N}_{1}}(\delta _{i}v_{i})+\left(k-
\abs{\mathcal{N}_{1}} \right)V^{\Delta }=1.
\label{equation: equilibrium-condition-6}
\end{equation}
To characterize an SSPE, it suffices to find $(\bm{x},\bm{v},\bm{p},\bm{\mu}
,V^{\Delta},V_L)$ that satisfies \eqref{equation: equilibrium-condition-1}-
\eqref{equation: equilibrium-condition-6}.

\paragraph{Equilibrium Existence}

Let $Y:= \sum_{i\in\mathcal{N}} \tilde{f}_{i}(x_{i})$. By 
\eqref{equation:
defintion-p}, we have $p_{i} = {\tilde{f}_{i}(x_{i})}/{Y}$, which implies
that 
\begin{equation}  \label{equation: equilibrium-condition-new-1}
x_{i} = \tilde{f}_{i}^{-1}(Yp_{i}) \text{, for } p_{i}\in\big[{\tilde{f}
_{i}(0)}/{Y},1\big],
\end{equation}
and 
\begin{equation}  \label{equation: equilibrium-condition-new-2}
\sum_{i\in\mathcal{N}}p_{i}=1.
\end{equation}

Substituting \eqref{equation: equilibrium-condition-new-1} into 
\eqref{equation: equilibrium-condition-1} yields 
\begin{equation}  \label{equation: equilibrium-condition-new-3}
\frac{Yc_{i}^{\prime}\left(\tilde{f}_{i}^{-1}(Yp_{i})\right)}{\tilde{f}
_{i}^{\prime }\left(\tilde{f}_{i}^{-1}(Yp_{i})\right)} \geq
(1-p_{i})(V_{L}+V^{\Delta}) - \mu_{i}V^{\Delta},\text{with equality holding
if }p_{i}>\frac{\tilde{f}_{i}(0)}{Y}.
\end{equation}
Rewriting \eqref{equation: equilibrium-condition-4} and 
\eqref{equation:
equilibrium-condition-5} and substituting 
\eqref{equation:
equilibrium-condition-3} into \eqref{equation: equilibrium-condition-6}
yield 
\begin{equation}  \label{equation: equilibrium-condition-new-4}
\mu_{i} = \frac{1}{V^{\Delta}}\text{med}\left\{0,V^{\Delta}(1-p_{i}), \frac{
V^{\Delta}}{\delta_{i}} - p_{i}(V_{L}+V^{\Delta})+ c_{i}\left(\tilde{f}
_{i}^{-1}(Yp_{i})\right)\right\},
\end{equation}
\begin{equation}  \label{equation: equilibrium-condition-new-5}
\sum_{i\in\mathcal{N}}\mu_{i} = k-1,
\end{equation}
and 
\begin{equation}  \label{equation: equilibrium-condition-new-6}
\sum_{i\in\mathcal{N}_{1}}\frac{\delta_{i}}{1-\delta_{i}}\left[
p_{i}V_{L}-c_{i}\left(\tilde{f}_{i}^{-1}(Yp_{i})\right)\right] + \left(k-
\abs{\mathcal{N}_{1}}\right)V^{\Delta}+V_{L} = 1,
\end{equation}
where $\text{med}\{\cdot,\cdot,\cdot\}$ gives the median of the input.

To prove equilibrium existence, it suffices to show that there exists $(
\bm{p},\bm{\mu},Y, V^{\Delta}, V_{L})$ to satisfy conditions 
\eqref{equation: equilibrium-condition-new-2}-
\eqref{equation:
equilibrium-condition-new-6}. The proof consists of four steps. First,
fixing $(Y, V^{\Delta}, V_{L})$, we show that there exists a unique $(\bm{p},
\bm{\mu})$ to satisfy \eqref{equation: equilibrium-condition-new-3} and 
\eqref{equation: equilibrium-condition-new-4}. Second, fixing $
(V^{\Delta},V_{L})$, there exists $Y\geq \sum_{i\in\mathcal{N}}\tilde{f}
_{i}(0)$ to satisfy \eqref{equation: equilibrium-condition-new-2}. Third,
fixing $V_{L}$, there exists $V^{\Delta}$ to satisfy 
\eqref{equation:
equilibrium-condition-new-5}. Last, we show that there exists $V_{L}$ to
satisfy \eqref{equation: equilibrium-condition-new-6}.

\paragraph{Step I}

Substituting \eqref{equation:
equilibrium-condition-new-4} into 
\eqref{equation:
equilibrium-condition-new-3} yields 
\begin{align}
\frac{Yc_{i}^{\prime}\big(\tilde{f}_{i}^{-1}(Yp_{i})\big)}{\tilde{f}
_{i}^{\prime }\big( \tilde{f}_{i}^{-1}(Yp_{i})\big)} \geq &\text{med}
\left\{(1-p_{i})(V_{L}+V^{\Delta}),(1-p_{i})V_{L}, V_{L} + V^{\Delta} - 
\frac{ V^{\Delta}}{\delta_{i}} - c_{i}(\tilde{f}_{i}^{-1}\big(Yp_{i})\big)
\right\},  \label{temp-proof-17}
\end{align}
with equality holding if $p_{i} > \frac{\tilde{f}_{i}(0)}{Y}$.

Let 
\begin{equation*}
\phi(p_{i}):= \frac{Yc_{i}^{\prime}\big(\tilde{f}_{i}^{-1}(Yp_{i})\big)}{
\tilde{f}_{i}^{\prime }( \tilde{f}_{i}^{-1}(Yp_{i}))}- \text{med}
\left\{(1-p_{i})(V_{L}+V^{\Delta}),(1-p_{i})V_{L}, V_{L} + V^{\Delta} - 
\frac{V^{\Delta}}{\delta_{i}} - c_{i}\big(\tilde{f}_{i}^{-1}(Yp_{i})\big)
\right\}.
\end{equation*}
By \Cref{Assumption: h_i} and \eqref{equation: tilde-f}, $\tilde{f}
_{i}(\cdot)$ is increasing and concave, which implies that $\phi(\cdot)$ strictly
increases with $p_{i}$. Therefore, if $\phi\big(\frac{\tilde{f}_{i}(0)}{Y}
\big) \geq 0$, or equivalently, if 
\begin{equation}  \label{temp-proof-4}
\frac{Yc_{i}^{\prime}(0)}{\tilde{f}_{i}^{\prime }(0)} \geq \text{med}
\left\{\left(1-\frac{\tilde{f}_{i}(0)}{Y}\right)V_{L}, \left(1-\frac{\tilde{f
}_{i}(0)}{Y}\right)(V_{L}+V^{\Delta}), V_{L}+V^{\Delta} - \frac{V^{\Delta}}{
\delta_{i}} \right\},
\end{equation}
then $p_{i} = \frac{\tilde{f}_{i}(0)}{Y}$. Otherwise, if $\phi\big(\frac{
\tilde{f}_{i}(0)}{Y}\big) < 0$, or equivalently, if 
\begin{equation}  \label{temp-proof-5}
\frac{Yc_{i}^{\prime}(0)}{\tilde{f}_{i}^{\prime }(0)} < \text{med}
\left\{\left(1-\frac{\tilde{f}_{i}(0)}{Y}\right)V_{L}, \left(1-\frac{\tilde{f
}_{i}(0)}{Y}\right)(V_{L}+V^{\Delta}), V_{L}+V^{\Delta} - \frac{V^{\Delta}}{
\delta_{i}} \right\},
\end{equation}
then $p_{i} > \frac{\tilde{f}_{i}(0)}{Y}$; moreover, $p_i$ is uniquely
pinned down by $\phi(p_{i}) = 0$, or equivalently, 
\begin{equation}  \label{temp-proof-6}
\frac{Yc_{i}^{\prime}\big(\tilde{f}_{i}^{-1}(Yp_{i})\big)}{\tilde{f}
_{i}^{\prime }( \tilde{f}_{i}^{-1}\big(Yp_{i})\big)} = \text{med}
\left\{(1-p_{i})(V_{L}+V^{\Delta}),(1-p_{i})V_{L}, V_{L} + V^{\Delta} - 
\frac{V^{\Delta}}{\delta_{i}} - c_{i}\big(\tilde{f}_{i}^{-1}(Yp_{i})\big)
\right\}.
\end{equation}
Further, $\mu_{i}$ can be uniquely solved from 
\eqref{equation:
equilibrium-condition-new-4}. Therefore, fixing $(Y, V^{\Delta}, V_{L})$,
there exists a unique pair $(p_i,\mu_i)$ to satisfy 
\eqref{equation:
equilibrium-condition-new-3} and 
\eqref{equation:
equilibrium-condition-new-4}, which we denote by $\big( p_{i}(Y, V^{\Delta},
V_{L}),\mu_{i}(Y, V^{\Delta}, V_{L})\big)$ with slight abuse of notation. 

\paragraph{Step II}

We show that fixing $(V^{\Delta}, V_{L})$ and $\big\{ p_{i}(Y, V^{\Delta},
V_{L}),\mu_{i}(Y, V^{\Delta}, V_{L})\big)\}_{i\in \mathcal{N}}$, there 
exists $Y\geq \sum_{i\in\mathcal{N}}\tilde{f}_{i}(0)$ to satisfy 
\eqref{equation: equilibrium-condition-new-2}. 

By definition of $p_{i}(Y, V^{\Delta}, V_{L})$, we have that $p_{i}(Y, V^{\Delta}, V_{L})
\geq \frac{\tilde{f}_{i}(0)}{Y}$, 
which implies 
\begin{equation*}
\sum_{i\in\mathcal{N}} p_{i}\left(\sum\nolimits_{j\in\mathcal{N}}\tilde{f}
_{j}(0), V^{\Delta}, V_{L}\right) \geq 1.
\end{equation*}
Next, we claim that 
\begin{equation*}
\lim\limits_{Y\to+\infty}\sum_{i\in\mathcal{N}} p_{i}(Y, V^{\Delta}, V_{L})
= 0.
\end{equation*}
To see this, first consider the case of $\tilde{f}_{i}^{\prime }(0) < +\infty$. Then \eqref{temp-proof-4}
holds as $Y$ approaches infinity, in which case $p_{i} = \frac{\tilde{f}
_{i}(0)}{Y}$ and 
\begin{equation*}
\lim\limits_{Y\to+\infty}p_{i}(Y, V^{\Delta}, V_{L}) =
\lim\limits_{Y\to+\infty}\frac{\tilde{f}_{i}(0)}{Y} = 0.
\end{equation*}
Next, consider the case of $\tilde{f}_{i}^{\prime }(0) = +\infty$. Then \eqref{temp-proof-5}
holds for all $Y$ and $p_{i}(Y, V^{\Delta},V_{L})$ solves 
\eqref{temp-proof-6}. As $Y $ approaches infinity, the right-hand side of 
\eqref{temp-proof-6} approaches infinity; therefore, the left-hand side must be
finite, indicating that $p_{i}(Y, V^{\Delta}, V_{L})$ approaches $0$.

By the intermediate value theorem, there exists $Y\geq \sum_{i\in\mathcal{N}
} \tilde{f}_{i}(0)$ such that 
\begin{equation*}
\sum_{i\in\mathcal{N}} p_{i}(Y, V^{\Delta}, V_{L}) = 1.
\end{equation*}
Fixing $(V^{\Delta},V_{L})$, we denote the largest $Y$ that solves the above equation by $Y(V^{\Delta},V_{L})$ in the subsequent analysis. 

\paragraph{Step III}

Fixing $V_{L}$, $Y(V^{\Delta},V_{L})$, and $\big\{ p_{i}(Y, V^{\Delta},
V_{L}),\mu_{i}(Y, V^{\Delta}, V_{L})\big)\}_{i\in \mathcal{N}}$, we show
that there exists $V^{\Delta} $ such that 
\eqref{equation:
equilibrium-condition-new-5} holds, i.e., 
\begin{equation}  \label{temp-proof-7}
\sum_{i\in\mathcal{N}} \mu_{i}\left(Y(V^{\Delta},V_{L}), V^{\Delta},
V_{L}\right) = k-1.
\end{equation}

First, consider the case in which $V^{\Delta}$ approaches $0$. For each $i\in
\mathcal{N}$, when $p_{i} = \frac{\tilde{f}_{i}(0)}{Y}$, we have that 
\begin{align*}
&\lim\limits_{V^{\Delta}\searrow 0}\mu_{i}\left(Y(V^{\Delta}, V_{L}),
V^{\Delta}, V_{L}\right) \\
=& \lim\limits_{V^{\Delta}\searrow 0}\frac{1}{V^{\Delta}}\text{med}
\left\{0, V^{\Delta}\left(1-\frac{\tilde{f}_{i}(0)}{Y(V^{\Delta},V_{L})}
\right), \frac{V^{\Delta}}{\delta_{i}} - \frac{\tilde{f}_{i}(0)}{
Y(V^{\Delta},V_{L})}(V_{L} + V^{\Delta}) \right\} =0,
\end{align*}
where the second equality follows from the fact that $\frac{V^{\Delta}}{
\delta_{i}} - \frac{\tilde{f}_{i}(0)}{Y(V^{\Delta},V_{L})}(V_{L} +
V^{\Delta}) \leq 0 \leq V^{\Delta}\big(1-\frac{\tilde{f}_{i}(0)}{
Y(V^{\Delta},V_{L})}\big)$ as $V^{\Delta }$ approaches $0$.

When $p_{i} > \frac{\tilde{f}_{i}(0)}{Y}$, by 
\eqref{equation:
equilibrium-condition-new-4}, $\mu_{i}\left(Y(V^{\Delta},V_{L}),V^{
\Delta},V_{L}\right) = 0$ for sufficiently small $V^{\Delta}$. 

Therefore, we have that 
\begin{equation*}
\lim\limits_{V^{\Delta}\searrow 0} \sum_{i\in\mathcal{N}}
\mu_{i}\left(Y(V^{\Delta},V_{L}), V^{\Delta}, V_{L}\right) = 0.
\end{equation*}

Next, consider the case in which $V^{\Delta}$ approaches infinity. For each $
i\in\mathcal{N}$, we have that 
\begin{align*}
0 \leq& V^{\Delta}\left[1-p_{i}\left(Y(V^{\Delta},V_{L}),V^{\Delta},V_{L}
\right)\right] \\
\leq& \frac{V^{\Delta}}{\delta_{i}}-p_{i}\left(Y(V^{\Delta},V_{L}),V^{
\Delta},V_{L}\right)(V_{L}+V^{\Delta})+c_{i}\left(\tilde{f}
_{i}^{-1}\left(Y(V^{\Delta},V_{L})p_{i}\left(Y(V^{\Delta},V_{L}),V^{
\Delta},V_{L}\right)\right)\right);
\end{align*}
together with \eqref{equation: equilibrium-condition-new-4}, we can obtain
that 
\begin{equation*}
\mu_{i}\left(Y(V^{\Delta},V_{L}), V^{\Delta}, V_{L}\right) =
1-p_{i}\left(Y(V^{\Delta},V_{L}),V^{\Delta},V_{L}\right) \text{, as }
V^{\Delta} \to +\infty.
\end{equation*}
Therefore, we have that 
\begin{equation*}
\lim\limits_{V^{\Delta}\to +\infty} \sum_{i\in\mathcal{N}}
\mu_{i}\left(Y(V^{\Delta},V_{L}), V^{\Delta}, V_{L}\right) =
\lim\limits_{V^{\Delta}\to +\infty} \sum_{i\in\mathcal{N}} \left[
1-p_{i}\left(Y(V^{\Delta},V_{L}), V^{\Delta}, V_{L}\right)\right] = n-1.
\end{equation*}
Note that $\mu_{i}(Y, V^{\Delta}, V_{L})$ and $Y(V^{\Delta},V_{L})$ are
continuous for all $i\in\mathcal{N}$, and $0\leq k-1 \leq n-1$. It follows immediately that
there exists $V^{\Delta}\geq 0$ to satisfy \eqref{temp-proof-7}. In what
follows, fixing $V_L$, let us denote the largest $V^{\Delta}$ that solves \eqref{temp-proof-7} by $
V^{\Delta}(V_{L})$.

\paragraph{Step IV}
We show that there exists $V_{L}\in [0,1]$ to satisfy 
\eqref{equation:
equilibrium-condition-new-6}, i.e., 
\begin{equation}  \label{temp-proof-8}
\sum_{i\in\mathcal{N}_{1}}\frac{\delta_{i}}{1-\delta_{i}}\left[
p_{i}V_{L}-c_{i}\left(\tilde{f}_{i}^{-1}(Yp_{i})\right)\right] + \left(k-
\abs{\mathcal{N}_{1}}\right)V^{\Delta} + V_{L} = 1,
\end{equation}
where $V^{\Delta} = V^{\Delta}(V_{L})$, $Y = Y(V^{\Delta}, V_{L})$, and $
p_{i} = p_{i}(Y, V^{\Delta}, V_{L})$ for $i\in\mathcal{N}$, as defined above.

Note that the left-hand side of \eqref{temp-proof-8} is always nonnegative; moreover, it is evident that the left-hand side is no less than $1$ when $V_{L} = 1$. To conclude the proof, it suffices to
show that $\lim\nolimits_{V_{L}\searrow 0 }V^{\Delta}(V_{L}) = 0$, from
which we can conclude that the left-hand side of \eqref{temp-proof-8}
approaches $0$ as $V_{L}\searrow 0$.

Suppose, to the contrary, that $\limsup\nolimits_{V_{L}\searrow 0
}V^{\Delta}(V_{L}) > 0$. It can then be verified that the following strict inequality holds as $V_{L} \searrow 0$ and $V^{\Delta} \to \limsup\nolimits_{V_{L}\searrow 0
}V^{\Delta}(V_{L})$:
\begin{equation*}
V^{\Delta }(1-p_{i}) < \frac{V^{\Delta}}{\delta_{i}} -
p_{i}(V_{L}+V^{\Delta}) +c_{i}\left(\tilde{f}_{i}^{-1}(Yp_{i})\right),\
\forall\, i\in\mathcal{N}.
\end{equation*}
Recall that $\mathcal{N}_{2}$ is nonempty by definition. That is, there
exists some agent $j\in\mathcal{N}_{2}$. By 
\eqref{equation:
equilibrium-condition-new-4}, we have that
\begin{equation*}
V^{\Delta }(1-p_{j}) \geq \frac{V^{\Delta}}{\delta_{j}} -
p_{j}(V_{L}+V^{\Delta}) +c_{j}\left(\tilde{f}_{j}^{-1}(Yp_{j})\right).
\end{equation*}
A contradiction.
\end{proof}

\medskip

\begin{description}
\item[Proof of \Cref{Theorem: optimal}] 
\end{description}

\begin{proof}
We first show that the optimum can be achieved by setting $k=1$. It suffices
to show that for each $(\bm{\alpha}, \bm{\beta}, k)$ and a resulting
equilibrium, there exists $(\bm{\hat{\alpha}}, \bm{\hat{\beta}})$ such that $
(\bm{\hat{\alpha}}, \bm{\hat{\beta}},1)$ induces the same equilibrium effort
profile $\bm{x}$ and recognition probabilities $\bm{p}$.

By \eqref{equation: defintion-p}, \eqref{equation: tilde-f}, and 
\eqref{equation: equilibrium-condition-1}, we have 
\begin{equation*}
p_{i} = \frac{\alpha_{i}f_{i}(x_{i}) + \beta_{i}}{\sum_{j\in\mathcal{N}}
\left[\alpha_{j}f_{j}(x_{j})+\beta_{j}\right]},
\end{equation*}
and 
\begin{equation}  \label{temp-proof-20}
c_{i}^{\prime}(x_{i})\frac{\alpha_{i}f_{i}(x_{i}) + \beta_{i}}{
\alpha_{i}f_{i}^{\prime }(x_{i})} \geq p_{i}(1-p_{i})\left(V_{L}+ \frac{
(1-p_{i}-\mu_{i})V^{\Delta}}{1-p_{i}} \right),
\end{equation}
with equality holding if $x_{i}>0$.

We construct $(\bm{\hat{\alpha}}, \bm{\hat{\beta}})$ as follows. For $x_{i}
= 0$, we set $(\hat{\alpha}_{i},\hat{\beta_{i}})= (0,p_{i})$. For $x_{i} > 0$
, note by \eqref{equation: equilibrium-condition-6} that we have that 
\begin{equation}  \label{temp-proof-21}
1 = \sum_{i\in\mathcal{N}_{1}}(\delta_{i}v_{i}) + (k-\abs{\mathcal{N}_{1}}
)V^{\Delta} + V_{L} \geq V^{\Delta } + V_{L},
\end{equation}
where the inequality follows from $v_{i}\geq 0$ and $\abs{\mathcal{N}_{1}}
\leq k-1$. Combining \eqref{temp-proof-20} and \eqref{temp-proof-21} yields 
\begin{equation*}
\frac{c_{i}^{\prime}(x_{i})f_{i}(x_{i})}{f_{i}^{\prime }(x_{i})} \leq
c_{i}^{\prime}(x_{i})\frac{\alpha_{i}f_{i}(x_{i}) + \beta_{i}}{
\alpha_{i}f_{i}^{\prime }(x_{i})} = p_{i}(1-p_{i})\left(V_{L}+ \frac{
(1-p_{i}-\mu_{i})V^{\Delta}}{1-p_{i}} \right) \leq p_{i}(1-p_{i}).
\end{equation*}
Define $\hat{\theta}_{i}:= p_{i}(1-p_{i}){f_{i}^{\prime }(x_{i})}/{
c_{i}^{\prime}(x_{i})} - f_{i}(x_{i})$. The above inequality indicates $\hat{
\theta}_{i}\geq 0$. Set 
\begin{equation}  \label{construction-alpha}
\left(\hat{\alpha}_{i},\hat{\beta}_{i}\right):= \left(\frac{p_{i}}{
f_{i}(x_{i})+ \hat{\theta}_{i}},\hat{\alpha}_{i}\hat{\theta}_{i}\right).
\end{equation}

It remains to verify that $(\bm{x},\bm{p})$ constitutes the unique equilibrium effort
profile and recognition probabilities under $(\bm{\hat{\alpha}},
\bm{\hat{\beta}},1)$. When $k=1$, the game degenerates to a standard static contest
with prize value of $1$. It suffices to show that the equilibrium recognition probability $p_i$ satisfies
\begin{equation}  \label{temp-proof-18}
p_{i} = \frac{\hat{\alpha}_{i}f_{i}(x_{i}) + \hat{\beta_{i}}}{\sum_{j\in
\mathcal{N}} \left[\hat{\alpha}_{j}f_{j}(x_{j})+\hat{\beta}_{j}\right]},
\end{equation}
and $x_{i}$ solves 
\begin{equation}  \label{temp-proof-19}
\max_{x_{i}\geq 0} \frac{\hat{\alpha}_{i}f_{i}(x_{i}) + \hat{\beta_{i}}}{
\sum_{j\in\mathcal{N}} \left[\hat{\alpha}_{j}f_{j}(x_{j})+\hat{\beta}_{j}
\right]} - c_{i}(x_{i}).
\end{equation}

Note that $p_{i} = \hat{\alpha}_{i}f_{i}(x_{i}) + \hat{\beta_{i}}$ for all $
i\in\mathcal{N}$ by construction (see, e.g., \eqref{construction-alpha}).
Therefore, $\sum_{j\in\mathcal{N}} (\hat{\alpha}_{j}f_{j}(x_{j})+\hat{\beta}
_{j}) = \sum_{j\in\mathcal{N}} p_{j} = 1$, which implies 
\eqref{temp-proof-18}. 

Next, we verify that $x_i$ solves the maximization problem 
\eqref{temp-proof-19}. For agent $i\in \mathcal{N}$ with $x_{i} = 0$, it is
evident that choosing $x_{i} = 0$ dominates $x_{i}>0$ under $(
\bm{\hat{\alpha}}, \bm{\hat{\beta}},1)$ because $\hat{\alpha}_{i} = 0$. For agent $
i\in \mathcal{N}$ with $x_{i} > 0$, by \eqref{construction-alpha}, we have that
\begin{equation*}
c_{i}^{\prime }(x_{i})\frac{\hat{\alpha}_{i}f_{i}(x_{i}) + \hat{\beta}_{i}}{
\hat{\alpha}_{i} f_{i}^{\prime}(x_{i}) } = c_{i}^{\prime}(x_{i}) \frac{
f_{i}(x_{i})+\hat{\theta}_{i}}{f_{i}^{\prime}(x_{i})} = p_{i}(1-p_{i}),
\end{equation*}
which is exactly the first-order condition for the maximization problem \eqref{temp-proof-19}.

The above analysis shows that the optimum can be achieved by $k=1$, in which
case the game reduces to a standard static contest. By Theorem $2$ in 
\citet{fu2020optimal}, the optimum can be achieved by choosing
multiplicative biases $\bm{\alpha}$ only and setting headstart $\bm{\beta}$
to zero under \Cref{Assumption: Objective function}.
\end{proof}

\begin{description}
	\item[Proof of \Cref{Thorem: monotonicity}] 
\end{description}

\begin{proof}
	It suffices to show that fixing an arbitrary equilibrium effort profile $\bm{x}^{\ast}$ and the resulting recognition probability profile $\bm{p}^{\ast}$ under some voting rule $k$ and contest rule $(\bm{\alpha},\bm{\beta})$, the designer can modify the contest rule under the less inclusive voting rule $k-1$ to induce the same equilibrium effort profile $\bm{x}^{\ast}$ and recognition probability profile $\bm{p}^{\ast}$.

	Plugging \eqref{equation: equilibrium-condition-4} into 
	\eqref{equation: equilibrium-condition-5}, we have that 
	\begin{equation}\label{temp-proof-k-30}
		\sum_{i\in\mathcal{N}_{2}} \left[\left(\frac{1}{\delta_{i}}
		-p_{i}^{\ast}\right)V^{\Delta}-p_{i}^{\ast}V_L+c_{i}(x_{i}^{\ast}) \right] +
		\sum_{i\in \mathcal{N}_{1}} (1-p_{i}^{\ast}) = (k-1)V^{\Delta}.
	\end{equation}
	Recall by \eqref{equation: equilibrium-condition-new-6}, we have that 
	\begin{equation}\label{temp-proof-k-31}
		\sum_{i\in \mathcal{N}_{1}} \frac{\delta_{i}}{1-\delta_{i}} \left[
		p_{i}^{\ast}V_{L}-c_{i}(x_{i}^{\ast}) \right] + \left(k-\vert\mathcal{N}
		_{1}\vert\right)V^{\Delta} + V_{L} = 1.
	\end{equation}

	Note that holding fixed $(\bm{x}^*,\bm{p}^*)$, we can adjust the contest rule to satisfy the above two equilibrium conditions as the voting rule $k$ varies, which gives a new pair $(V^\Delta,V_L)$. To prove the theorem, it remains to verify the following first-order condition under the less inclusive voting rule $k-1$ and the new pair $(V^\Delta,V_L)$: 
	\begin{equation*}
		\frac{c_{i}^{\prime}(x_{i}^{\ast})f_{i}(x_{i}^{\ast})}{f_{i}^{
				\prime}(x_{i}^{\ast})} \leq p_{i}^{\ast}(1-p_{i}^{\ast}) \left[V_{L} +
		V^{\Delta} - \frac{\mu_{i}}{1-p_{i}^{\ast}}V^{\Delta} \right], \forall \, i\in\mathcal{N}.
	\end{equation*}
	Evidently, it suffices to show that the effective prize spread, $V_{L} +
	V^{\Delta} - \frac{\mu_{i}}{1-p_{i}^{\ast}}V^{\Delta}$, is non-increasing in 
	$k$.

	We treat $k$ as a continuous variable. Clearly, $\mu_{i}$, $V^{\Delta}$, and $V_{L}$ are all continuous in $k$. Moreover, for all but finitely many values of $k$, the sets $\mathcal{N}_{1}$, $\mathcal{N}_{2}$, and $\mathcal{N}_{3}$ remain unchanged in a neighborhood of $k$, which indicates that $\mu_{i}$, $V^{\Delta}$, and $V_{L}$ are differentiable with respect to $k$.  Therefore, it suffices to show that the derivative of the effective prize spread, $V_{L} + V^{\Delta} - \frac{\mu_{i}}{1-p_{i}^{\ast}}V^{\Delta}$, with respect to $k$ is nonpositive whenever it is differentiable. Taking the derivatives of \eqref{temp-proof-k-30} and \eqref{temp-proof-k-31} with respect to $k$ yields that
	
	\begin{equation*}
		\frac{dV_{L}}{dk} =- \frac{(\mathcal{B}+\mathcal{D})V^{\Delta}}{\mathcal{A}
			\mathcal{D} + \mathcal{B}\mathcal{C}}\text{ and } \frac{dV^{\Delta}}{dk} = - \frac{(
			\mathcal{C}-\mathcal{A})V^{\Delta}}{\mathcal{A}\mathcal{D} + \mathcal{B}
			\mathcal{C}},
	\end{equation*}
	where 
	\begin{align*}
		\mathcal{A} :=& 1 + \sum_{i\in \mathcal{N}_{1}} \frac{p_{i}^{\ast}\delta_{i}}{
			1-\delta_{i}} > 0, \\
		\mathcal{B} :=&k-\vert\mathcal{N}_{1}\vert > 0, \\
		\mathcal{C} :=&\sum_{i\in\mathcal{N}_{2}} p_{i}^{\ast} > 0, \\
		\mathcal{D} :=& \sum_{i\in \mathcal{N}_{1}} (1-p_{i}^{\ast}) + \sum_{i\in
			\mathcal{N}_{2}}\left(\frac{1}{\delta_{i}}-p_{i}^{\ast}\right) - (k-1) > 0.
	\end{align*}
	Evidently, $\mathcal{A}\mathcal{D}+\mathcal{B}\mathcal{C}>0$ and $\mathcal{B}
	+ \mathcal{D} > 0$. Moreover, we have that 
	\begin{equation*}
		\mathcal{C} - \mathcal{A} = \sum_{i\in\mathcal{N}_{2}} p_{i}^{\ast} - 1 -
		\sum_{i\in \mathcal{N}_{1}} \frac{p_{i}^{\ast}\delta_{i}}{1-\delta_{i}} \leq
		- \sum_{i\in \mathcal{N}_{1}} \frac{p_{i}^{\ast}\delta_{i}}{1-\delta_{i}} \leq
		0.
	\end{equation*}
	Therefore, we have that
	\begin{equation}  \label{temp-proof-29}
		\frac{dV_{L}}{dk} =- \frac{(\mathcal{B}+\mathcal{D})V^{\Delta}}{\mathcal{A}
			\mathcal{D} + \mathcal{B}\mathcal{C}} \leq 0 \text{ and } \frac{dV^{\Delta}}{dk} = - 
		\frac{(\mathcal{C}-\mathcal{A})V^{\Delta}}{\mathcal{A}\mathcal{D} + \mathcal{
				B}\mathcal{C}} \geq 0.
	\end{equation}
	
	We consider the following three cases.
	
	\paragraph{Case I: $\bm{i\in\mathcal{N}_{1}}$.} By \eqref{equation: equilibrium-condition-4}, $\mu_{i}=1-p_{i}^{\ast}$ and the
	effective prize spread is $V_{L}$. We can then conclude from \eqref{temp-proof-29} that $
	V_{L}$ is non-increasing in $k$.
	
	\paragraph{Case II: $\bm{i\in\mathcal{N}_{2}}$.} By \eqref{equation: equilibrium-condition-4}, the effective prize spread is 
	\begin{align*}
		V_{L} + V^{\Delta} - \frac{\mu_{i}}{1-p_{i}^{\ast}}V^{\Delta} =& V_{L} +
		V^{\Delta} - \frac{V^{\Delta}}{\delta_{i}(1-p_{i}^{\ast})} + \frac{
			p_{i}^{\ast}}{1-p_{i}^{\ast}}(V_{L}+V^{\Delta}) - c_{i}(x_{i}^{\ast}) \\
		=& \frac{V_{L}}{1-p_{i}^{\ast}} - \frac{(1-\delta_{i})V^{\Delta}}{
			\delta_{i}(1-p_{i}^{\ast})} - c_{i}(x_{i}^{\ast}).
	\end{align*}
	Carrying out the algebra, we can obtain that
	\begin{equation*}
		\frac{d}{dk}\left(V_{L} + V^{\Delta} - \frac{\mu_{i}}{1-p_{i}^{\ast}}
		V^{\Delta} \right) = \frac{1}{1-p_{i}^{\ast}}\times \frac{dV_{L}}{dk} - \frac{1-\delta_{i}}{\delta_{i}(1-p_{i}^{\ast})}\times \frac{dV^{\Delta}}{dk} \leq 0.
	\end{equation*}
	
	\paragraph{Case III: $\bm{i\in\mathcal{N}_{3}}$.} By \eqref{equation: equilibrium-condition-4}, $\mu_{i}=0$ and the
	effective prize spread is $V_{L}+V^{\Delta}$. By \eqref{temp-proof-29}, we
	have that 
	\begin{equation*}
		\frac{d(V_{L}+V^{\Delta})}{dk} = \frac{(\mathcal{B}+\mathcal{C}+\mathcal{D}-
			\mathcal{A})V^{\Delta}}{\mathcal{A}\mathcal{D} + \mathcal{B}\mathcal{C}}.
	\end{equation*}
	It remains to prove
	\begin{equation*}
		\mathcal{B}+\mathcal{C}+\mathcal{D}-\mathcal{A} \geq 0.
	\end{equation*}
	Carrying out the algebra, we have that 
	\begin{align*}
		\mathcal{B}+\mathcal{C}+\mathcal{D}-\mathcal{A} =& k-\vert\mathcal{N}
		_{1}\vert + \sum_{i\in\mathcal{N}_{2}} p_{i}^{\ast} + \sum_{i\in \mathcal{N}
			_{1}} (1-p_{i}^{\ast}) + \sum_{i\in\mathcal{N}_{2}}\left(\frac{1}{\delta_{i}}
		-p_{i}^{\ast}\right) - (k-1) - 1 - \sum_{i\in \mathcal{N}_{1}} \frac{
			p_{i}^{\ast}\delta_{i}}{1-\delta_{i}} \\
		=& \sum_{i\in\mathcal{N}_{2}} \frac{1}{\delta_{i}} - \sum_{i\in \mathcal{N}
			_{1}} \frac{p_{i}^{\ast}}{1-\delta_{i}} \\
		\geq& 2\vert\mathcal{N}_{2}\vert - 2\sum_{i\in \mathcal{N}_{1}} p_{i}^{\ast}
		\geq 0,
	\end{align*}
	where the first inequality follows from $\delta_{i} \leq 
	\frac{1}{2}$. This concludes the proof.
\end{proof}

\newpage
\section*{Appendix B: Derivation for Examples}

\begin{description}
\item[Derivation for Equilibria in \Cref{Example: nonmonotone in k}] 
\end{description}

First, consider the case of $k=1$. The game reduces to a static Tullock
contest. Let $Y:=\sum_{i\in \mathcal{N}}\eta_{i}x_{i}$. The equilibrium conditions can be derived as 
\begin{equation*}
Y = 1 - p_{i},
\end{equation*}
from which we can solve for the equilibrium aggregate effort $Y$, the equilibrium
recognition probabilities $\bm{p}=(p_1,p_2,p_3,p_4)$, and the equilibrium efforts $
\bm{x}=(x_1,x_2,x_3,x_4)$ as follows: 
\begin{equation*}
Y = \frac{3}{4},
\end{equation*}
\begin{equation*}
p_{i} = 1 - Y = \frac{1}{4},\ \forall i\in\mathcal{N},
\end{equation*}
and 
\begin{equation*}
\bm{x} = Y \bm{p}\oslash\bm{\eta} = \left(\frac{3}{16},\frac{15}{16},\frac{15}{16},\frac{15}{16}\right).
\end{equation*}

Next, consider the case of $k=2$. The equilibrium conditions in the proof of 
\Cref{Theorem: equilibrium}---i.e., conditions 
\eqref{equation:
equilibrium-condition-new-1}, \eqref{equation: equilibrium-condition-new-2}, 
\eqref{equation: equilibrium-condition-new-3}, 
\eqref{equation:
equilibrium-condition-new-4}, \eqref{equation: equilibrium-condition-new-5},
and \eqref{equation: equilibrium-condition-new-6}---for this example can be expressed as follows: 
\begin{equation*}
Yp_{i} = \eta_{i}x_{i},
\end{equation*}
\begin{equation*}
\sum_{i\in\mathcal{N}}p_{i} = 1,
\end{equation*}
\begin{equation*}
Y = (1-p_{i})(V_{L}+V^{\Delta}) - \mu_{i}V^{\Delta},
\end{equation*}
\begin{equation*}
\mu_{i} = \frac{1}{\delta_{i}} - p_{i} - \frac{p_{i}V_{L}-Yp_{i}}{
V^{\Delta}},
\end{equation*}
\begin{equation*}
\sum_{i\in\mathcal{N}}\mu_{i} = 1,
\end{equation*}
\begin{equation*}
\frac{1}{9}\left[p_{1}V_{L} - Yp_{1}\right] + V_{L} + V^{\Delta} = 1.
\end{equation*}
It can be verified that $\bm{p} = (0.2322,0.2559, 0.2559,0.2559)$, $
\bm{x} = (0.1711,0.9433,0.9433,0.9433)$, $\bm{\mu} = (0.7678,
0.0774,0.0774,0.0774)$, $V_{L} = 0.9600$, and $V^{\Delta} = 
0.0342$ constitute an SSPE of the game. The equilibria for the cases of 
$k=3$ and $k=4$ can be similarly verified.

\medskip

\begin{description}
\item[Derivation for the Optimal Recognition Mechanism in \Cref{example: optimal_beta>0}] 
\end{description}

We demonstrate the optimality of $(\bm{\alpha}^{\ast },\bm{\beta}^{\ast })$
in \Cref{example: optimal_beta>0}. When the designer sufficiently cares about
the profile of agents' recognition probabilities---i.e., when $\lambda \gg
1/c$---the optimal equilibrium winning probability profile must be $\bm{p} =
(\frac{1}{3 },\frac{1}{3},\frac{1}{3})$ and the designer's payoff at $\bm{p}
= (\frac{1}{3},\frac{1}{3},\frac{1}{3})$ reduces to $\Lambda = x_{1} + x_{2} + x_{3}$
. When $c$ is sufficiently small, agent 3 is excessively strong and the
designer's payoff is mainly determined by $x_3$. Therefore, it suffices to
show that $(\bm{\alpha}^{\ast}, \bm{\beta}^{\ast})$ maximizes $x_{3}$ among
all rules $(\bm{\alpha},\bm{\beta})$ that induce $\bm{p} = (\frac{1}{3},
\frac{1}{3},\frac{1}{3})$.

Fix $\bm{p} = (\frac{1}{3},\frac{1}{3},\frac{1}{3})$. We first rewrite the
equilibrium conditions in the proof of \Cref{Theorem: equilibrium}---i.e.,
conditions \eqref{equation: equilibrium-condition-new-1}-
\eqref{equation:
equilibrium-condition-new-6}. 

Evidently, condition \eqref{equation: equilibrium-condition-new-2} is
satisfied and condition \eqref{equation: equilibrium-condition-new-1}
becomes
\begin{equation}  \label{temp-proof-22}
\alpha_{i}^{\ast}x_{i} + \beta_{i}^{\ast } = \frac{Y}{3},\ \forall\,
i\in\{1,2,3\}.
\end{equation}
Next, consider condition \eqref{equation: equilibrium-condition-new-3}. The
condition holds with equality for $x_i>0$. Further, if $x_{i}=0$ for some $
i\in\mathcal{N}$ and the strict inequality holds, we can increase $
\alpha_{i} $ until the equality holds and at the same time keep unchanged the
equilibrium effort profile $\bm{x}$ and recognition probabilities $\bm{p}$. Therefore, we can assume that equality holds for all agents and
the condition becomes 
\begin{equation}  \label{temp-proof-23}
\frac{Yc_{i}}{\alpha_{i}^{\ast}} = \frac{2(V_{L}+V^{\Delta}) }{3} -
\mu_{i}V^{\Delta },\ \forall\, i\in\{1,2,3\}.
\end{equation}
Substituting \eqref{temp-proof-23} into \eqref{temp-proof-22} yields 
\begin{equation}  \label{temp-proof-10}
3c_{i}x_{i} \leq \frac{2(V_{L}+V^{\Delta}) }{3} - \mu_{i}V^{\Delta },\
\forall\, i\in\{1,2,3\},
\end{equation}
with equality holding if $\beta_{i}^{\ast } = 0$. To establish the
optimality of headstarts, it suffices to show that the inequality is strict
for at least one agent.

Conditions \eqref{equation: equilibrium-condition-new-4}, 
\eqref{equation:
equilibrium-condition-new-5}, and 
\eqref{equation:
		equilibrium-condition-new-6} are 
\begin{equation}  \label{temp-proof-11}
\mu_{i} = \left\{ 
\begin{array}{ll}
\frac{2}{3} \leq \frac{1}{\delta_{i}} - \frac{1}{3} - \frac{V_{L}}{
3V^{\Delta}} + \frac{ c_{i}x_{i}}{V^{\Delta}}, & i\in\mathcal{N}_{1}, \\ 
\frac{1}{\delta_{i}} - \frac{1}{3} - \frac{V_{L}}{3V^{\Delta}} + \frac{
c_{i}x_{i}}{V^{\Delta}} \in \left[0,\frac{2}{3}\right], & i\in\mathcal{N}
_{2}, \\ 
0 \geq \frac{1}{\delta_{i}} - \frac{1}{3} - \frac{V_{L}}{3V^{\Delta}} + 
\frac{ c_{i}x_{i}}{V^{\Delta}}, & i\in\mathcal{N}_{3},
\end{array}
\right.
\end{equation}
\begin{equation}  \label{temp-proof-12}
\mu_{1}+\mu_{2}+\mu_{3}=1,
\end{equation}
and 
\begin{equation}  \label{temp-proof-24}
\sum_{i\in\mathcal{N}_{1}}\frac{\delta_{i}}{1-\delta_{i}}\left(\frac{V_{L}}{3
} -c_{i}x_{i}\right) + \left(2-\abs{\mathcal{N}_{1}}\right)V^{\Delta} +V_{L} = 1.
\end{equation}
Substituting \eqref{temp-proof-11} in \eqref{temp-proof-10} yields that
\begin{equation}  \label{temp-proof-13}
c_{i}x_{i} \leq \left\{ 
\begin{array}{ll}
\frac{2}{9}V_{L}, & i\in\mathcal{N}_{1}, \\ 
\frac{1}{4}\left[V_{L}-(\frac{1}{\delta_{i}}-1)V^{\Delta} \right], & i\in 
\mathcal{N}_{2}, \\ 
\frac{2}{9}(V_{L}+V^{\Delta}), & i\in\mathcal{N}_{3},
\end{array}
\right.
\end{equation}
from which we can conclude $c_{i}x_{i} \leq \frac{2V_{L}}{9}$ for $i\in
\mathcal{N}_{1}$; together with \eqref{temp-proof-24}, we can obtain that 
\begin{equation}  \label{temp-proof-14}
\sum_{i\in\mathcal{N}_{1}}\frac{\delta_{i}}{1-\delta_{i}}\times\frac{V_{L}}{
9 } + \left(2-\abs{\mathcal{N}_{1}}\right)V^{\Delta} +V_{L} \leq 1.
\end{equation}

In what follows, we will show that $c_{3}x_{3} \leq \frac{30}{144-\delta_{3}}
$, and the equality holds if and only if $\bm{\alpha}^{\ast } = (\frac{62Y}{
35}, \frac{ 62Y}{37}, \frac{62Yc}{39})$ and $\bm{\beta}^{\ast } = (0, \frac{
17Y}{222}, 0 )$. Consider the following three cases.

\paragraph{Case I: $\bm{3\in\mathcal{N}_{1}}$.}

Note that $\abs{\mathcal{N}_{1}}\leq k-1 = 1$, we have that $\mathcal{N}_{1}
= \{3\}$. By \eqref{temp-proof-14}, we can obtain that 
\begin{equation*}
\left[1 + \frac{\delta_{3}}{9(1-\delta_{3})} \right]V_{L} + V^{\Delta} \leq
1;
\end{equation*}
together with \eqref{temp-proof-24}, we can obtain that 
\begin{equation*}
c_{3}x_{3} \leq \frac{2V_{L}}{9} \leq \frac{2(1-\delta_{3})}{9-8\delta_{3}}< 
\frac{30}{144-\delta_{3}}.
\end{equation*}

\paragraph{Case II: $\bm{3\in\mathcal{N}_{2}}$.}

By \eqref{temp-proof-11} and \eqref{temp-proof-13}, we have that 
\begin{equation*}
0 \leq \frac{1}{\delta_{3}} - \frac{1}{3} - \frac{V_{L}}{3V^{\Delta}} + 
\frac{c_{3}x_{3}}{V^{\Delta}} \leq \frac{1}{\delta_{3}} - \frac{1}{3} - 
\frac{V_{L}}{3V^{\Delta}} + \frac{V_{L}-(\frac{1}{\delta_{3}}-1)V^{\Delta}}{
4V^{\Delta}}.
\end{equation*}
Carrying out the algebra, we can obtain that 
\begin{equation}  \label{temp-proof-15}
V_{L} \leq \left(\frac{9}{\delta_{3}}-1\right)V^{\Delta} = \frac{35}{4}
V^{\Delta}.
\end{equation}
Further, $3\notin\mathcal{N}_{1}$ implies that $\mathcal{N} _{1} \in
\left\{\{1\},\{2\},\emptyset\right\}$, and thus \eqref{temp-proof-14}
becomes 
\begin{align}  \label{temp-proof-16}
1 \geq& \left\{ 
\begin{array}{ll}
\frac{16}{15}V_{L} + V^{\Delta}, & \text{if } \mathcal{N}_{1} = \{1\} \\ 
\frac{10}{9}V_{L} + V^{\Delta}, & \text{if } \mathcal{N}_{1} = \{2\} \\ 
V_{L} + 2V^{\Delta}, & \text{if } \mathcal{N}_{1} = \emptyset
\end{array}
\right\} \geq \frac{16}{15}V_{L} + V^{\Delta},
\end{align}
where the last inequality follows from \eqref{temp-proof-15}.

Combining \eqref{temp-proof-13}, \eqref{temp-proof-15}, and 
\eqref{temp-proof-16}, we have that 
\begin{equation*}
c_{3}x_{3} \leq \frac{1}{4}\left[V_{L}-(\frac{1}{\delta_{3}}-1)V^{\Delta} 
\right] \leq \frac{30}{144-\delta_{3}} = \frac{13}{62}.
\end{equation*}
Note that equality holds in condition \eqref{temp-proof-13} if and only if $
\beta_{3}^{\ast } = 0$. Further, equality holds in condition 
\eqref{temp-proof-15} only if $\mu_{3} = 0$. Last, equality holds in
condition \eqref{temp-proof-16} if and only if $\mathcal{N}_{1} = \{1\}$ and 
$\beta_{1}^{\ast } = 0$.

Because $\mathcal{N}_{1} = \{1\}$ and $\mu_{3} = 0$, we have that $\mu_{1} = 
\frac{2}{3}$ from \eqref{temp-proof-11}; together with \eqref{temp-proof-12}
, we have $\mu_{2} = \frac{1}{3}$. Moreover, by \eqref{temp-proof-11}, we
can conclude $2\in \mathcal{N}_{2}$, which implies that $\mathcal{N}_{2} =
\{2,3\}$ and $\mathcal{N}_{3} = \emptyset$.

Combining \eqref{temp-proof-15} and \eqref{temp-proof-16} (recall that
equality holds in these two conditions), we can obtain $V_{L} = \frac{105 }{124}$
and $V^{\Delta} = \frac{3}{31}$; together with \eqref{temp-proof-13}, we
have $x_{1} = \frac{2V_{L}}{9} = \frac{35}{186}$. Substituting $\mu_{2} = 
\frac{1}{3}$, $V_{L} = \frac{105}{124}$ and $V^{\Delta} = \frac{3}{31}$ in \eqref{temp-proof-11}, we can obtain that $x_{2} = \frac{V_{L}-4V^{\Delta}}{3
} =\frac{19}{124}$.

Last, we solve for $(\bm{\alpha}^{\ast },\bm{\beta}^{\ast })$. Recall that $
\beta_{i}^{\ast } = 0$ for $i\in\{1,3\}$. Therefore, $\alpha_{i}^{\ast } = 
\frac{Y }{3x_{i}}$ from \eqref{temp-proof-22}. For $i=2$, we have $x_{2} = 
\frac{19}{124}$. Further, by \eqref{temp-proof-23}, we have $\frac{Y}{
\alpha_{2}^{\ast }} = \frac{2V_{L}+V^{\Delta}}{3} = \frac{37}{62}$, which
implies that $\alpha_{2}^{\ast } = \frac{62Y}{37}$; together with 
\eqref{temp-proof-22}, we can conclude $\beta_{2}^{\ast } = \frac{Y}{3}
-\alpha_{2}^{\ast }x_{2} = \frac{17Y}{222}$.

In summary, the equality holds in $c_{3}x_{3} \leq \frac{30}{144-\delta_{3}}$
if and only if $\bm{\alpha}^{\ast } = (\frac{ 62Y}{35}, \frac{62Y}{37}, 
\frac{62Yc}{39})$ and $\bm{\beta}^{\ast } = (0, \frac{17Y}{222}, 0 )$, under
which the equilibrium is $\bm{x}= (\frac{35}{186}, \frac{19}{124}, \frac{39}{
186c} )$, $\bm{p} = (\frac{1}{3},\frac{1}{3}, \frac{1}{3})$, $\bm{\mu}= (
\frac{2}{3},\frac{1}{3},0)$, $V_{L} = \frac{105}{ 124}$, and $V^{\Delta} = 
\frac{3}{31}$.

\paragraph{Case III: $\bm{3\in\mathcal{N}_{3}}$.}

Condition \eqref{temp-proof-11}, together with the postulated $3\in\mathcal{N
}_{3}$, implies that $\mu_{3} = 0$. Analogous to derivation of 
\eqref{temp-proof-15}, we can obtain that 
\begin{equation}  \label{temp-proof-25}
V_{L} > \left(\frac{9}{\delta_{3}}-1\right)V^{\Delta} = \frac{35}{4}
V^{\Delta}.
\end{equation}

Suppose $\mathcal{N}_{1}\not=\emptyset$. By \eqref{temp-proof-16}, we have
that 
\begin{equation}  \label{temp-proof-26}
1 \geq \frac{16V_{L}}{15} + V^{\Delta }.
\end{equation}
Combining \eqref{temp-proof-10}, \eqref{temp-proof-25}, and 
\eqref{temp-proof-26} yields that 
\begin{equation*}
c_{3}x_{3} \leq \frac{2(V_{L}+V^{\Delta})}{9} < \frac{30}{144-\delta_{3}}.
\end{equation*}

Next, suppose $\mathcal{N}_{1}=\emptyset $; together with $3\in \mathcal{N}
_{3}$ and $k=2$, we can conclude $\mathcal{N}_{2}=\{1,2\}$. It follows from 
\eqref{temp-proof-24} that 
\begin{equation}
V_{L}+2V^{\Delta }=1.  \label{temp-proof-27}
\end{equation}
Recall $\mu _{3}=0$. Combining \eqref{temp-proof-11}, \eqref{temp-proof-12},
and \eqref{temp-proof-13}, we can obtain that 
\begin{equation*}
1=\mu _{1}+\mu _{2}=\frac{1}{\delta _{1}}+\frac{1}{\delta _{2}}-\frac{2}{3}-
\frac{2V_{L}}{3V^{\Delta }}+\frac{x_{1}+x_{2}}{V^{\Delta }}\leq 4-\frac{
2V_{L}}{3V^{\Delta }}+\frac{V_{L}}{2V^{\Delta }}-\frac{2}{3},
\end{equation*}
which in turn implies that 
\begin{equation}
V_{L}\leq 14V^{\Delta }.  \label{temp-proof-28}
\end{equation}
Therefore, 
\begin{equation*}
c_{3}x_{3}\leq \frac{2(V_{L}+V^{\Delta })}{9}\leq \frac{5}{24}<\frac{30}{
144-\delta _{3}},
\end{equation*}
where the first inequality follows from \eqref{temp-proof-10} and the second
inequality from \eqref{temp-proof-27} and \eqref{temp-proof-28}.

\medskip
\begin{description}
	\item[Derivation for the Optimal Voting Rule in \Cref{Example: non-monotonicity}] 
\end{description}

By \Cref{Theorem: optimal}, the optimum can be achieved by setting $k=1$. It remains to show that the optimum can be achieved by setting $k=5$, but not $k=4$. Evidently, when $\lambda$ is sufficiently large, the optimum requires
that $\bm{p} = \bm{\tilde{p}}$. Moreover, when $\gamma$ is sufficiently
large, each agent $i$'s equilibrium effort $x_{i}$ cannot exceed $\tilde{x
}_{i}$. Therefore, it suffices to show that fixing $\bm{p}=\bm{\tilde{p}}$,
the equilibrium effort is $\bm{x} = \bm{\tilde{x}}$ at $k=5$, and the
designer cannot induce $\bm{x} = \bm{\tilde{x}}$ at $k=4$.

When $k=5$, it can be verified that $\mathcal{N}
_{1} = \{1,2,3\}$, $\mathcal{N}_{2} = \{4,5,6\}$, and $\mathcal{N}_{3} =
\{7\}$. Moreover, the equilibrium effort is $\bm{\tilde{x}}$, and
equilibrium winning probability is $\bm{\tilde{p}}$. In this case, agent $7$
's effective prize spread is $V_{L} + V^{\Delta} = 0.8399$, and his first-order
condition holds with equality: 
\begin{equation*}
r\tilde{x}_{7} c_{7}^{\prime}(\tilde{x}_{7}) = (V_L+V^{\Delta})\tilde{p}
_{7}(1-\tilde{p}_{7}).
\end{equation*}

Next, we show that when $k=4$, the designer cannot induce $\bm{p}=
\bm{\tilde{p}}$ and $\bm{x}=\bm{\tilde{x}}$ simultaneously. In fact, fixing $
\bm{p}=\bm{\tilde{p}}$ and $\bm{x}=\bm{\tilde{x}}$, by 
\eqref{equation:
equilibrium-condition-new-4}-\eqref{equation: equilibrium-condition-new-6},
we have that $V_L = 0.7439$ and $V^{\Delta} = 0.0669$, with $V_L+V^\Delta <0.8399$
. However, agent $7$'s first-order condition requires that 
\begin{equation*}
r\tilde{x}_{7}^{r} = \tilde{x}_{7} c_{7}^{\prime}(\tilde{x}_{7}) = 0.8399\times \tilde{p}_{7}(1-\tilde{p}_{7}) \leq (V_L+V^{\Delta})\tilde{p}_{7}(1-\tilde{p}_{7}).
\end{equation*}
A contradiction.

\end{document}